\documentclass[acmsmall,screen,nonacm]{acmart}

\AtBeginDocument{%
  \providecommand\BibTeX{{%
      \normalfont B\kern-0.5em{\scshape i\kern-0.25em b}\kern-0.8em\TeX}}}

\setcopyright{acmcopyright}
\copyrightyear{2022}
\acmYear{2022}
\acmDOI{10.1145/1122445.1122456}

\usepackage{listings}
\usepackage{xcolor}

\def \mcorefontsize {\normalsize}

\def \mcorefont {\fontfamily{pcr}\selectfont\mcorefontsize}

\definecolor{ckeywords}{rgb}{0.13,0.13,1}
\definecolor{ccomments}{rgb}{0,0.5,0.5}
\definecolor{cstrings}{rgb}{0,0.5,0}
\definecolor{cwarnings}{rgb}{1,0.5,0}

\lstdefinelanguage{MCore}{
    morekeywords={Lam,con,else,end,fix,if,in,lam,lang,let,match,recursive,sem,syn,then,type,use,utest,with,hole,switch,case,independent},
    otherkeywords={->},
    keywordstyle=\color{ckeywords},
    morekeywords=[2]{mexpr,include,never},
    keywordstyle=[2]\color{cwarnings},
    morecomment=[l][\color{ccomments}]{--},
    morestring=[b]",
    stringstyle=\color{cstrings},
    sensitive=true,
    basicstyle=\mcorefont,
    breaklines=true,
    escapeinside={--*}{*--},
    numbers=left,
    stepnumber=1,
    numberstyle=\tiny,
    xleftmargin=0pt, % numbers will be in the margins
    columns=fixed, % same width for all characters
    showstringspaces=false,
    mathescape=true,
    breaklines=true,
    breakatwhitespace=true,
    mathescape=true,
    showstringspaces=false
  }

\lstnewenvironment{mcore}
  {
    \lstset{
      language=MCore,
      numbers=none,
      xleftmargin=0pt
    }
  }{}

\lstnewenvironment{mcore-lines}
  {
    \lstset{
      language=MCore,
      xleftmargin=0pt
    }
  }{}

\newcommand{\mcoreinline}[1]{\lstinline[{language=MCore}]|#1|}

\def\mcorefontsize{\footnotesize}

\usepackage{listings}
\usepackage{wrapfig, blindtext}
\usepackage{tikz}
\usetikzlibrary{snakes,arrows,shapes,automata}

\newcommand{\ignore}[1]{}

\usepackage{fancyvrb}

\usepackage[section]{algorithm}
\usepackage[noend]{algpseudocode}

\makeatletter
\def\algbackskip{\hskip-\ALG@thistlm}
\makeatother

\usepackage{xcolor}
\usepackage{pdfcomment}

\usepackage{xspace}
\usepackage{amsmath}
\theoremstyle{definition}

\numberwithin{alg}{section}

\usepackage[inline]{enumitem}

\usepackage{tikz}
\usepackage{pgfplots}

\newcommand{\RangeInclusive}[2]{[#1,#2]}
\newcommand{\Union}[0]{\bigcup}

\usepackage{multirow}
\usepackage{enumitem}

\usepackage{subcaption}

\newcommand{\MeasSet}[0]{M}

\newcommand{\HolesSet}[0]{H}

\begin{document}

%%
%% The "title" command has an optional parameter,
%% allowing the author to define a "short title" to be used in page headers.

\title{Programming with Context-Sensitive Holes using Dependency-Aware Tuning}

% Ingredients for updated title:
% Dependency-Aware Tuning
% Automatic Dependency Detection
% Static Dependency Analysis

%%
%% The "author" command and its associated commands are used to define
%% the authors and their affiliations.
%% Of note is the shared affiliation of the first two authors, and the
%% "authornote" and "authornotemark" commands
%% used to denote shared contribution to the research.
\author{Linnea Stjerna}
\email{lstjerna@kth.se}
\orcid{0000-0002-9578-5407}
\affiliation{
  \institution{EECS and Digital Futures, KTH Royal Institute of Technology}
  \city{Stockholm}
  \country{Sweden}
}
\author{David Broman}
\email{dbro@kth.se}
\orcid{0000-0001-8457-4105}
\affiliation{
  \institution{EECS and Digital Futures, KTH Royal Institute of Technology}
  \city{Stockholm}
  \country{Sweden}
}

%%
%% By default, the full list of authors will be used in the page
%% headers. Often, this list is too long, and will overlap
%% other information printed in the page headers. This command allows
%% the author to define a more concise list
%% of authors' names for this purpose.
\renewcommand{\shortauthors}{Trovato and Tobin, et al.}

%%
%% The abstract is a short summary of the work to be presented in the
%% article.
\begin{abstract}
  Developing efficient and maintainable software systems is both hard and time
  consuming. In particular, non-functional performance requirements involve many
  design and implementation decisions that can be difficult to take early during
  system development. Choices---such as selection of data structures or where
  and how to parallelize code---typically require extensive manual tuning that
  is both time consuming and error-prone. Although various auto-tuning
  approaches exist, they are either specialized for certain domains or require
  extensive code rewriting to work for different contexts in the code. In this
  paper, we introduce a new methodology for writing programs with holes, that
  is, decision variables explicitly stated in the program code that enable
  developers to postpone decisions during development. We introduce and evaluate
  two novel ideas: (i) \emph{context-sensitive holes} that are expanded by the
  compiler into sets of decision variables for automatic tuning, and (ii)
  \emph{dependency-aware tuning}, where static analysis reduces the search space
  by finding the set of decision variables that can be tuned
  \emph{independently} of each other. We evaluate the two new concepts in a
  system called Miking, where we show how the general methodology can be used
  for automatic algorithm selection, data structure decisions, and
  parallelization choices.
\end{abstract}

\maketitle

\section{Introduction}
\label{sec:intro}

Software developers constantly face implementation choices that affect
performance, such as choices of data structures, algorithms, and parameter
values. Unfortunately, traditional programming languages lack support for
expressing such alternatives directly in the code, forcing the programmer to commit to
certain design choices up front.
To improve the performance of a program, a developer can
profile the code and manually tune the program by explicitly executing
the program repeatedly, testing and changing different algorithm
choices and parameter settings. However, such manual tuning is both tedious and
error prone because the number of program alternatives grows
exponentially with the number of choices. Besides choices of program
parameters, hardware properties, such cache layout and core
configurations make manual optimization even more challenging.

An attractive alternative to manual tuning of programs is to perform
the tuning automatically.  Conceptually, a tuning problem is an
optimization problem where the search space consists of potential
program alternatives and the goal is to minimize an objective, such
as execution time, memory usage, or code size. The search space is
explored using some search technique, which can be performed either
offline (at compile-time) or online (at runtime).
In this work, we focus on offline tuning.

Several offline auto-tuning tools have been developed to target
problems in specific domains. For instance, ATLAS~\cite{atlas} for
linear algebra, FFTW~\cite{fftw} and SPIRAL~\cite{spiral-overview-18}
for linear digital processing, PetaBricks~\cite{petabricks-09} for
algorithmic choice, as well as tools for choosing sorting algorithm
automatically~\cite{sorting-04}. Although these domain-specific tuning
approaches have shown to work well in their specific area,
they are inherently targeting a specific domain and cannot be used
in general for other kinds of applications.

In contrast to domain-specific tuning, \emph{generic} auto-tuners offer
solutions that work across domains and may be applicable to arbitrary software
systems. For instance, approaches such as ATF~\cite{atf},
OpenTuner~\cite{open-tuner}, CLTune~\cite{cltune},
HyperMapper~\cite{hypermapper}, and PbO~\cite{pbo} make it possible for
developers to specify unknown variables and possible values. The goal of the
tuner is then to assign values to these variables optimally, to minimize for
instance execution time. In particular, existing generic auto-tuners assign
values to variables \emph{globally}, i.e., each unknown variable is assigned the
same value throughout the program.

In this work, we focus on two problems with state-of-the-art
methods. Firstly, only using global decision variables does not scale in
larger software projects and it violates fundamental software
engineering principles. Specifically, a global decision variable does
not take into consideration \emph{the context} from where a function
may be called. Global decision variables can of course be manually
added in all calling contexts, but such manual refactoring of code
makes code updates brittle and harder to maintain.
Secondly, auto tuning is typically computationally expensive, since the search
space consists of all combinations of decision variable values. This is because
decision variables are dependent on each other in general. However, if a subset
of the decision variables is \emph{independent}, then the auto tuner wastes
precious time exploring unnecessary configurations.

In this paper, we introduce a methodology where a software developer postpones
design decisions during code development by explicitly stating \emph{holes} in
the program. A hole is a decision variable for the auto tuner, with a given
default value and domain, such as integer range or Boolean type. In contrast to
existing work, we introduce \emph{context-sensitive holes}. This means that a
hole that is specified in a program (called a base hole) can be expanded into a
set of decision variables (called context holes) that take the calling context
into consideration.
The compiler statically analyses the call graph of the program and transforms
the program so that the calling context is maintained during runtime. Only paths
through the call graph up to a certain length are considered, to avoid a
combinatorial explosion in the number of variables.

In our approach, context-sensitive holes can be automatically tuned using input
data, without manual involvement.
We develop and apply a novel static \emph{dependency analysis} to build a
bipartite dependency graph, which encodes the dependencies among the
context-sensitive holes. In contrast to existing approaches, the dependency
analysis is automatic, though optional annotations may be added to increase the
accuracy. The dependency graph is used to reduce the search space in a method we
call \emph{dependency-aware tuning}. Using this method, the auto tuner
concentrates its computation time on exploring only the necessary configurations
of hole values.

Specifically, we make the following contributions:
\begin{itemize}
\item We propose a general methodology for programming with
  context-sensitive holes that enables the software developer to
  postpone decisions to a later stage of automatic tuning. In
  particular, we discuss how the methodology can be used for algorithm
  and parallelization decisions
  (Section~\ref{sec:holes}).
\item To enable the proposed methodology, we propose a number of
  algorithms for performing program transformations for efficiently
  maintaining calling context during runtime
  (Section~\ref{sec:transformations}).
\item We design and develop a novel static dependency analysis, and use the
  result during tuning to reduce the search space of the problem
  (Sections~\ref{sec:dependency-analysis}--\ref{sec:dep-tuning}).
\item We design and develop a complete solution for the proposed idea
  within the Miking framework~\cite{Broman:2019}, a software framework for developing
  general purpose and domain-specific languages and compilers
  (Section~\ref{sec:implementation}).
\end{itemize}

\noindent To evaluate the approach, we perform experiments on several
non-trivial case studies that are not originally designed for the approach
(Section~\ref{sec:eval}).

\section{Programming with Context-Sensitive Holes}\label{sec:holes}

In this section, we first introduce the main idea behind the methodology of
programming with holes. We then discuss various programming examples, using
global and context-sensitive holes.

\subsection{Main Idea}\label{sec:main-idea}

Traditionally, an idea is implemented into a program by making a number of
implementation choices, and gradually reducing the design space.
By contrast, when programming with holes, we delay the decision of
implementation choices. Design choices---such as what parts of the
program to parallelize or which algorithms to execute in different
circumstances---are instead left unspecified by specifying holes, that
is, decision variables stated directly in the program code. Instead of
taking the design decision up-front, an auto-tuner framework makes use
of input data and runtime profiling information, to make decisions of
filling in the holes with the best available alternatives. Thus, the
postponed decisions can be automated and based on more information,
resulting in less ad hoc and more informed decisions.

\subsection{Global Holes}
\label{sec:global-holes}
The key concept in our methodology is the notation of \emph{holes}.
A hole is an unknown variable whose optimal value is to be decided by the
auto-tuner, defined in a program using the keyword \mcoreinline{hole}. The
\mcoreinline{hole} takes as arguments the type of the hole (either
\mcoreinline{Boolean} or \mcoreinline{IntRange}) and its default value.
Additionally, the \mcoreinline{IntRange} type expects a minimum and maximum
value.

\begin{example}\label{ex:global}
  To illustrate the idea of a hole, we first give a small example illustrating
  how to choose sorting algorithms based on input data length. Consider the
  following program, implemented in the Miking core
  language\footnote{We use the Miking core language in the rest of the paper
  because the experimental evaluation is implemented in Miking. The concepts
  and ideas presented in the paper are, however, not bound to any specific
  language or runtime system.}:  %
\begin{lstlisting}[language=MCore,firstline=6,lastline=10]
let sort = lam seq.
  let threshold = hole (IntRange             --*\label{l:sort1}*--
    {default = 10, min = 0, max = 10000}) in --*\label{l:sort2}*--
  if leqi (length seq) threshold then insertionSort seq  --*\label{l:sort-ite-1}*--
  else mergeSort seq  --*\label{l:sort-ite-2}*--
\end{lstlisting}
  \sloppypar{%
    The example defines a function \mcoreinline{sort} using the
    \mcoreinline{let} construction. The function has one parameter
    \mcoreinline{seq}, defined using an anonymous lambda function
    \mcoreinline{lam}.
    Lines~\ref{l:sort1}--\ref{l:sort2} define a hole with possible values in
    the range $\RangeInclusive{0}{10000}$ and default value~$10$. The program
    then chooses to use \verb|insertionSort| if the length of the sequence is
    less than the threshold, and \verb|mergeSort| otherwise. That is, an
    auto-tuner can use offline profiling to find the optimal threshold value
    that gives the overall best result. Note also that the default value can be
    used if the program is compiled without the tuning stage.} \qed{}

\end{example}

The example above illustrates the use of a \emph{global hole}, i.e., a
hole whose value is chosen globally in the program. Although global
holes are useful, they do not take into consideration the calling
context.

\subsection{Context-Sensitive Holes}
\label{sec:context-sensitive-holes}
All holes that are explicitly stated in the program code using the
\mcoreinline{hole} syntax represent \emph{base holes}. As illustrated
in the previous example, a base hole that does not take into
consideration the calling context is the same thing as a global
hole. One of the novel ideas in this paper is the concept of
\emph{context-sensitive} holes.  In contrast to a global hole, the
value of a context-sensitive hole varies depending on which
call path is taken to the place in the program where the hole is
defined. They are useful in programs where we believe the optimal
value of the hole varies depending on the context. Context-sensitive
holes are implicitly defined from a base hole, taking into
consideration the different possible call paths reaching the base
hole. The idea is illustrated with the following example.

\begin{example}\label{example-map1}
  Consider the higher-order function \mcoreinline{map}, which applies a
  function~\mcoreinline{f} to all elements in a sequence~\mcoreinline{s}. The function can
  be applied either sequentially or in parallel (given that~\mcoreinline{f} is
  side-effect free).
  The optimal choice between the sequential or parallel version likely depends
  partly on the length of the sequence, but also on the nature of the
  function~\mcoreinline{f}.
  In a large program, a common function such as \mcoreinline{map} is probably called
  from many different places in the program, with varying functions \mcoreinline{f}
  and sequences. Therefore, globally deciding whether to use the sequential or
  parallel version may result in suboptimal performance. \qed
\end{example}

To define a context-sensitive hole, we provide an additional (optional)
field~\mcoreinline{depth}, which represents the length of the most recent function
call history that should influence the choice of the value of the hole.

\begin{example}\label{example-map2}
  The following is an implementation of \mcoreinline{map} that chooses between the
  parallel and sequential versions (\mcoreinline{pmap} and \mcoreinline{smap},
  respectively).
\begin{lstlisting}[language=MCore,firstline=5,lastline=7]
let map = lam f. lam s.
  let par = hole (Boolean {default = false, depth = 1}) in --*\label{l:bool-hole2}*--
  if par then pmap f s else smap f s  --*\label{l:par-ite}*--
\end{lstlisting}
  Line~\ref{l:bool-hole2} defines a Boolean hole \mcoreinline{par} with default
  value \mcoreinline{false} (no parallelization). The \mcoreinline{depth = 1}
  tells the tuner that the value of \mcoreinline{par} should consider the call
  path one step backward. That is, two calls to \mcoreinline{map} from different
  call locations can result in different values of \mcoreinline{par}.\qed{}
\end{example}

\begin{example}\label{example-map3}
  Given that we choose the parallel map \mcoreinline{pmap} in
  Example~\ref{example-map2}, another choice is how many parts to split the
  sequence into, before mapping~\mcoreinline{f} to each part in parallel. We can
  implement this choice with an integer range hole that represents the
  \mcoreinline{chunkSize}, i.e. the size of each individual part.
\begin{lstlisting}[language=MCore,firstline=25,lastline=31]
let pmap = lam f. lam s.
  let chunkSize = hole (IntRange
    {default = 10000, depth = 2, min = 1, max = 1000000000}) in
  let chunks = split s chunkSize in
  let tasks = smap (lam chunk.
    async (lam. smap f chunk)) chunks in
  join (smap await tasks)
\end{lstlisting}
  Function \mcoreinline{split} splits the sequence \mcoreinline{s} into chunks of
  size (at most) \mcoreinline{chunkSize}, that \mcoreinline{async} sends the tasks to a
  thread pool, and that \mcoreinline{await} waits for the tasks to be finished.

  Note the choice of \mcoreinline{depth = 2} in this example. We expect the function
  \mcoreinline{pmap} to be called from \mcoreinline{map}, which has \mcoreinline{depth = 1}.
  Since we want to capture the context from \mcoreinline{map} we need to increment
  the depth parameter by one.\qed
\end{example}

The choice of the \mcoreinline{depth} parameter for a hole should intuitively be
based on the number of function calls backward that might influence the choice
of the value of the hole. A larger depth might give a better end result, but it
also gives a larger search space for the automatic tuner, which influences the
tuning time.

Note that in order to use global holes to encode the same semantics as the
context-sensitive holes, we need to modify the function signatures.
For example, in Example~\ref{example-map2}, which has a hole of depth $1$, we
can let \mcoreinline{map} take \mcoreinline{par} as a parameter, and pass a
global hole as argument each time we call \mcoreinline{map}.
For holes of depth $d > 1$, this strategy becomes even more cumbersome, as each
function along a call path need to pass on global holes as arguments.
By instead letting the compiler handle the analysis of contexts, we do not need
to modify function signatures, and we can easily experiment with the depth
parameter. Additionally, holes can be ``hidden'' inside libraries, so that a
user does not need to be aware of where the holes are defined, but still benefit
from context-sensitive tuning.
Another advantage of context-sensitive holes compared to global holes is that
the compiler may use the knowledge that two context holes originate from the
same base hole to speed up the tuning stage.

\section{Program Transformations}\label{sec:transformations}

This section covers the program transformations necessary for maintaining the
context of each hole during runtime of the program.
The aim of the program transformations is that the resulting program maintain
contexts at a minimum runtime overhead.
Sections~\ref{sec:context-intuition} and~\ref{sec:graph-coloring} provide
definitions and a conceptual illustration of contexts, while
Section~\ref{sec:impl-transformations} covers the implementation in more detail.

\subsection{Definitions}
\label{sec:context-intuition}

\newcommand{\Depth}[0]{\delta}
\newcommand{\Home}[0]{\eta}
\newcommand{\SinglePaths}[0]{\Sigma}
\newcommand{\CallStrings}[0]{\mathit{CS}}
\newcommand{\Program}[0]{p}

Central in our discussion of contexts are call graphs. Given a program
$\Program{}$ with holes, its \emph{call graph} is a quintuple $G = (V, E, L, S,
H)$, where:
\begin{itemize}
\item the set of vertices $V$ represents the set of functions in $\Program{}$,
\item each edge $e\in E$ is a triple $e = (v_1, v_2, l)$ that represents a
  function call in $\Program{}$ from $v_1\in V$ to $v_2\in V$, labeled with
  $l\in L$,
\item $L$ is a set of labels uniquely identifying each call site in $\Program$;
  that is: $|E| = |L|$,
\item $S \subseteq V$ is the set of entry points in the program,
\item the triple $H = (n, \Depth, \Home)$ contains the number $n$ of base holes;
  a function $\Depth : \RangeInclusive{1}{n} \rightarrow \mathbb{N}$, which maps
  each base hole to its depth parameter; and a function $\Home :
  \RangeInclusive{1}{n} \rightarrow V$, which maps each base hole to the vertex
  $v\in V$ in which the hole is defined.
\end{itemize}

Furthermore, a \emph{call string} in a call graph is a string from the alphabet
$L$, describing a path in the graph from a start vertex $v_s\in S$ to some end
vertex $v_e\in V$. Let $\CallStrings_i$ denote the set of call strings of the
$i$th hole, that is, the set of call strings starting in some vertex $v_s\in S$
and ending in $\Home(i) \in V$.

\begin{wrapfigure}{r}{0.5\linewidth}
  \centering
  \includegraphics[width=0.5\textwidth,trim={1cm 0cm 0cm 0cm},clip,scale=0.6]{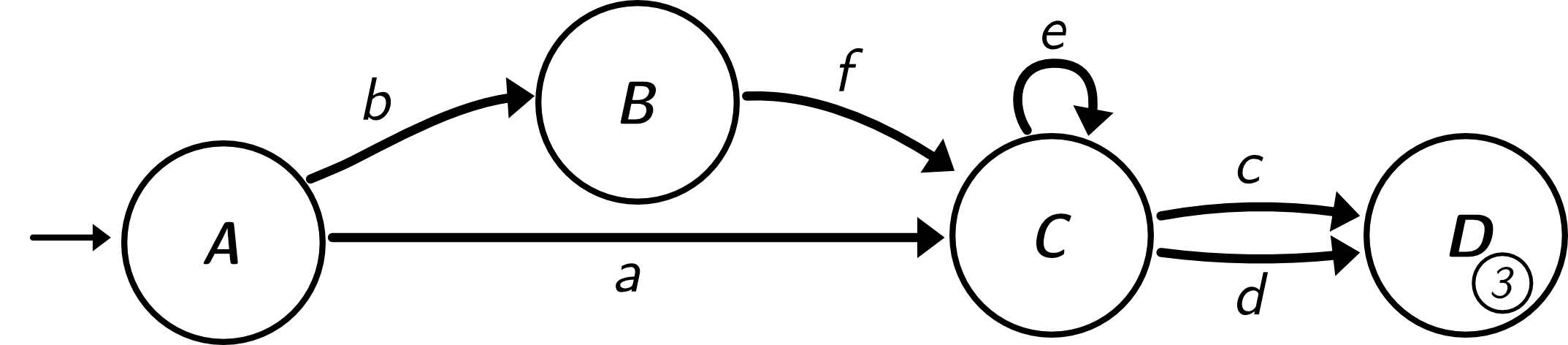}
  \caption{Call graph representing a program with four functions: $A$, $B$,
    $C$, and $D$. Each edge represents a function call from a certain call
    site in the program, and is marked with a unique
    label.}\label{fig:call-graph}
\end{wrapfigure}

\begin{example}\label{ex:call-graph-1}
  The call graph in Figure~\ref{fig:call-graph} is the quintuple:
\begin{alignat*}{2}
  (&\{A,B,C,D\},\\
   &\{(A,B,b), (A,C,a), (B,C,f), (C,C,e), (C,D,c), (C,D,d)\},\\
   &\{a,b,c,d,e,f\},\\
   &\{A\},\\
   &(1, \{1 \mapsto 3\}, \{1 \mapsto D\}))
 \end{alignat*}
 In Figure~\ref{fig:call-graph}, note that we mark a base hole with its depth as
 a smaller circle within the vertex where it is defined, that we mark each entry
 point of the graphs with an incoming arrow, and that there are two function
 calls from $C$ to $D$, hence two edges with different labels. The set
 $\CallStrings_1$ of call strings from $A$ to $D$ in Figure~\ref{fig:call-graph}
 is: $$ \mathit{ae^*c}, \mathit{ae^*d}, \mathit{bfe^*c}, \mathit{bfe^*d} $$ where
 the Kleene-star ($^*$) denotes zero or more repetitions of the previous label.
 \qed
\end{example}

Call strings provide the context that is relevant for context-sensitive holes.
Ideally, the tuning should find the optimal value for each call string in
$\CallStrings_i$ leading to the $i$th hole. However, in
Example~\ref{ex:call-graph-1} we see that there can be infinitely many call
strings, which means that we would have infinitely many decision variables to
tune. Therefore, we wish to partition the call strings into equivalence classes,
and tune each equivalence class separately.
%
% The number of equivalence classes for a base holes defines the number of
%
This leads to the question of how to choose the equivalence relation.

One possible choice of equivalence relation is to consider two call strings that
have an equal suffix (that is, an equal ending) as being equal. We can let the
length of the suffix be $d$, where $d$ is the context depth of the hole:
\begin{example}\label{ex:call-strings-1}
  Let $\sim_i$ be an equivalence relation defined on each set $\CallStrings_i$,
  such that:
  $$
  s_1 \sim_i s_2 \ \text{iff} \ \text{suffix}_{\Depth{i}}(s_1) =
  \text{suffix}_{\Depth{i}}(s_2)\text{, } s_1, s_2 \in \CallStrings_i
  $$
  where $\text{suffix}_d(s)$ returns the last $d$ labels of a call string $s$,
  or all of the labels if the length of the string is less than $d$.
  We choose a canonical representation from each equivalence class as the result
  from the $\text{suffix}_d$ function.
  The call strings from Example~\ref{ex:call-graph-1} have the following
  canonical representations (that is, unique results after applying
  $\text{suffix}_3$ to the call strings):
  $$
    \mathit{ac}
    , \mathit{aec}
    , \mathit{eec}
    , \mathit{ad}
    , \mathit{aed}
    , \mathit{eed}
    , \mathit{bfc}
    , \mathit{fec}
    , \mathit{bfd}
    , \mathit{fed}
  $$
  Note that the canonical representations are \emph{suffixes} of call strings
  but not always call strings by strict definition, as they do not always start
  in a start vertex $v_s\in S$. We call these canonical representations
  \emph{context strings}.
  \qed
\end{example}
While the equivalence relation in Example~\ref{ex:call-strings-1} at least gives
an upper bound on the number of decision variables to tune, it may still result
in a large number of equivalence classes. Limiting the number of recursive calls
that are considered results in a more coarse-grained partitioning:
\begin{example}\label{ex:call-strings-2}
  Consider the equivalence relation $\sim_i^r$, which is like $\sim_i$ from
  Example~\ref{ex:call-strings-1}, but where we consider at most $r$ repetitions
  of any label, for some parameter $r$. That is, if a string contains more than
  $r$ repetitions of a label, then we keep the $r$ rightmost occurrences. 
  For example, using $r = 1$ in the call strings of
  Example~\ref{ex:call-graph-1} yields the $8$ context strings:
  $$
  \mathit{ac}
  , \mathit{aec}
  %, \mathit{eec}
  , \mathit{ad}
  , \mathit{aed}
  %, \mathit{eed}
  , \mathit{bfc}
  , \mathit{fec}
  , \mathit{bfd}
  , \mathit{fed}
  $$
  Note that compared to the context strings in Example~\ref{ex:call-strings-1},
  we have filtered out $\mathit{eec}$ and $\mathit{eed}$ as they include more
  than $1$ repetition of the label $e$.
  For example, this means that the two call strings $\mathit{aec}$ and
  $\mathit{aeec}$ belongs to the same equivalence class, namely the class
  represented by the context string $\mathit{aec}$.
  \qed
\end{example}

Using the definition of context strings, we can finally define what we mean by
context-sensitive holes. Given some equivalence relation $\sim$, a base hole is
expanded (by the compiler) into $c$ number of \emph{context} holes, where $c$ is
the number of equivalence classes (i.e., the number of context strings) under
the relation $\sim$.
If $\Depth(i) = 0$, then the hole is \emph{global}, and $c = 1$.
In Example~\ref{ex:call-strings-1}, there are $10$~context holes, and in
Example~\ref{ex:call-strings-2} there are $8$.
Thus, the choice of the equivalence relation influences the number of decision
variables to tune: few equivalence classes give fewer variables to tune and
potentially a shorter tuning time, while more classes might increase the tuning
time, but give better performance of the resulting program.

\subsection{Graph Coloring for Tracking Contexts}
\label{sec:graph-coloring}

Each time a context-sensitive hole is used, we need to decide which equivalence
class the current call string belongs to, in order to know which value of the
hole to use.
The challenge is to introduce tracking and categorization of call strings in the
program with minimum runtime overhead.
A naive approach is to explicitly maintain the call string during runtime.
However, this requires book keeping of auxiliary data structures and potentially
makes it expensive to decide the current context string.
This section describes an efficient graph coloring scheme that leaves a color
trail in the call graph, thereby implicitly maintaining the call history during
runtime of the program.
We first discuss the underlying equivalence relation that the method implements
(Section~\ref{sec:eq-rel}), and then divide our discussion of the graph coloring
into two parts: complete programs (Section~\ref{sec:complete-program}) and
separately compiled library code (Section~\ref{sec:library}).

\subsubsection{Equivalence Relation}
\label{sec:eq-rel}

\newcommand{\Concat}[0]{\oplus}

The equivalence relation that the graph coloring method implements is an
approximation of the $\sim_i^1$ relation of Example~\ref{ex:call-strings-2}. The
difference is that we do not track the call string beyond a recursive (including
mutually recursive) call.
This is because a recursive call overwrites the call history in graph coloring,
as we will see in Section~\ref{sec:graph-coloring}.
The context strings for the call strings from the previous section are:
$$
\mathit{ac}
, \mathit{ec}
, \mathit{ad}
, \mathit{ed}
, \mathit{bfc}
, \mathit{bfd}
$$
Compared to the context strings in Example~\ref{ex:call-strings-2}, the strings
$\mathit{aed}$ and $\mathit{fed}$ are merged into $\mathit{ed}$, and the
strings $\mathit{aec}$ and $\mathit{fec}$ are merged into $\mathit{ec}$.
A consequence is that, for instance, the call strings $\mathit{bfed}$ and
$\mathit{aed}$ belong to the same equivalence class, namely the class
represented by $\mathit{ed}$.

Algorithm~\ref{algo:context-strings} describes how to explicitly compute the
context strings for the $i$th hole. The recursive sub-procedure
$\textsc{ContextStringsDFS}$ traverses the graph in a backwards depth-first
search manner. It maintains the current vertex $v$ (initially $\Home(i)$), the
current string $s$ (initially the empty string $\epsilon$), the set of visited
vertices $U$ (initially $\varnothing$) and the remaining depth $d$ (initially
$\Depth(i)$).
\begin{algorithm}
  \caption{Algorithm for computing the context strings of a
    hole.}\label{algo:context-strings}
  \textbf{Input} Call graph $G = (V,E,L,S,H)$, index $i$ of the base hole. \\
  \noindent\textbf{Output} Set of context strings of the hole. \\
  \begin{algorithmic}[1]
    \Procedure{ContextStrings}{$G$,$i$}
    \Procedure{ContextStringsDFS}{$v$, $s$, $U$, $d$}
    \If{$d=0 \lor \text{inc}(G,v) = \varnothing \lor v\in U$}
      \Return $\{s\}$\label{l:cs-return}
    \Else
    \State
    $\mathit{CS} \gets
      \bigcup\limits_{(v_p,v,l) \in \text{inc}(G,v)}
      \textsc{ContextStringsDFS}(v_p, s \Concat l, U \Union \{v\}, d-1)$\label{l:cs-rec}
      \If{$v\in S$}
      \Return $\mathit{CS} \Union \{s\}$\label{l:cs-ret-1}
      \Else \textbf{ return} $\mathit{CS}$\label{l:cs-ret-2}
      \EndIf
    \EndIf
    \EndProcedure
    \State \Return \textsc{ContextStringsDFS}($\Home(i)$, $\epsilon$, $\varnothing$, $\Depth(i)$)
    \EndProcedure
  \end{algorithmic}
\end{algorithm}
Line~\ref{l:cs-return} returns a singleton set if the depth is exhausted, if the
set of incoming edges to $v$ is empty, or if $v$ is visited. The function
$\text{inc}(G,v)$ returns the set of incoming edges to vertex $v$ in $G$.
Line~\ref{l:cs-rec} recursively computes the context strings of the preceding
vertices of $v$, and takes the union of the results. The $\Concat$ operator adds
a label to a string.
Lines~\ref{l:cs-ret-1}--\ref{l:cs-ret-2} return the final result. If the current
vertex $v$ is a start vertex, then the current string $s$ is a context string
starting in $v$, and is therefore added to the result. Otherwise, we return the
result of the recursive calls.

\subsubsection{Coloring a Complete Program}
\label{sec:complete-program}

\newcommand{\Setup}[0]{Setup}
\newcommand{\PruneGraph}[0]{Setup}
\newcommand{\TraverseEdge}[0]{TraverseEdge}
\newcommand{\Categorize}[0]{Categorize}
\newcommand{\ColorV}[0]{c_V}
\newcommand{\ColorE}[0]{c_L}

In the case of a complete program, the program has a single entry point, for
instance, a main function where the execution starts. We will now walk through a number of examples, showing how graph coloring works conceptually.

\begin{figure}[b!]
  \begin{subfigure}[t]{.45\textwidth}
  \centering
    \includegraphics[width=0.9\textwidth]{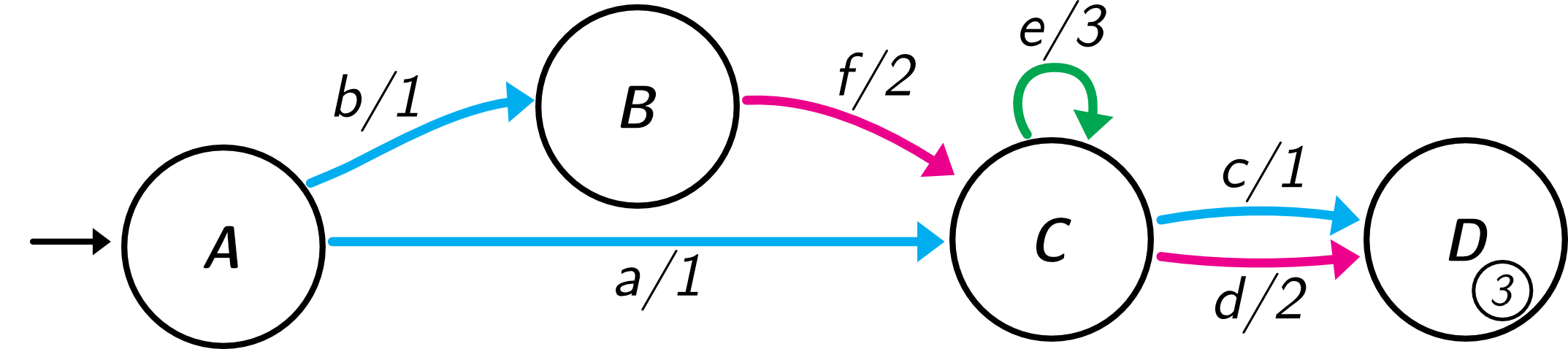}
    \caption{Initial state of the call graph when performing graph coloring:
      all vertices are white, and each edge is colored so that the colors of the
      incoming edges for each node are different.
      For readability on a black and white rendering of the figure, we mark each
      edge with an integer representing the color in addition to coloring the
      edge ($1$ for blue, $2$ for magenta, and $3$ for
      green).}\label{fig:color-init}
  \end{subfigure}\hfill
  \begin{subfigure}[t]{.45\textwidth}
  \centering
    \includegraphics[width=0.9\textwidth]{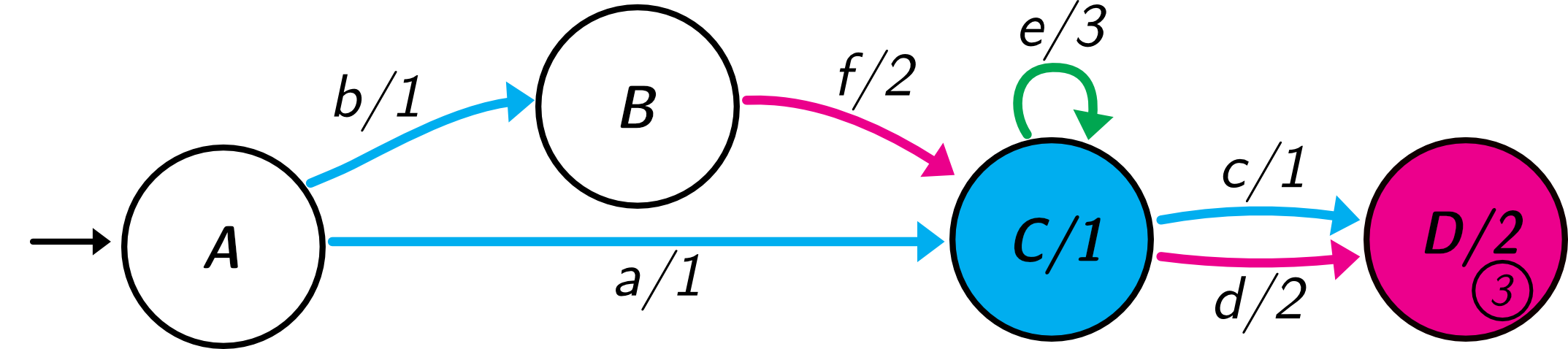}
    \caption{State of the call graph after the call string $\mathit{ad}$ is
      taken. For readability, for each non-white node we write the integer
      representing the color in the vertex in addition to coloring
      it.}\label{fig:color-step-1}
  \end{subfigure}
  \begin{subfigure}[t]{.45\textwidth}
  \centering
    \includegraphics[width=0.9\textwidth]{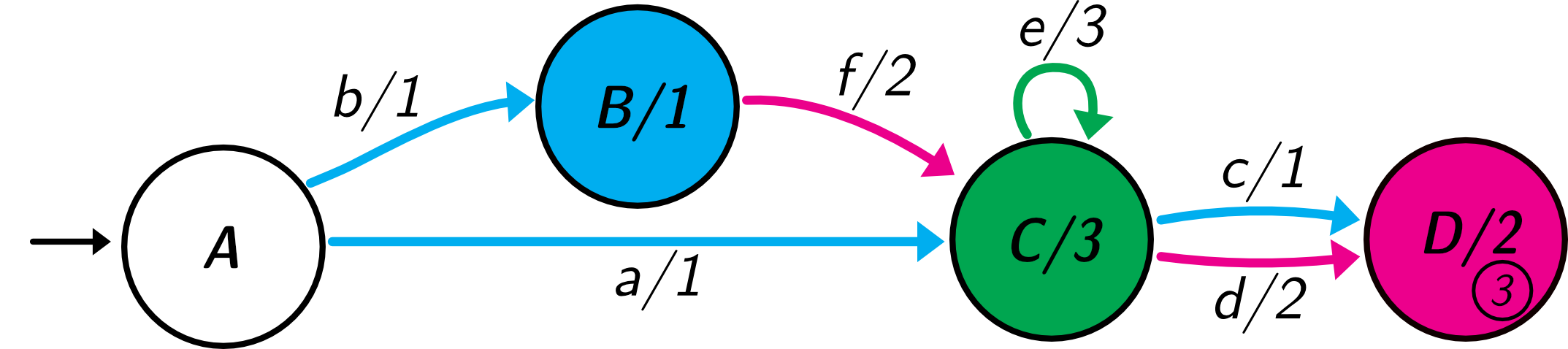}
    \caption{State of the call graph after the call string $\mathit{bfed}$ is
      taken. Because of the recursive call in $C$, we cannot trace the calls
      further backward from $B$. The current context string is therefore
      $ed$.}\label{fig:color-step-2}
  \end{subfigure}\hfill
  \begin{subfigure}[t]{.45\textwidth}
  \centering
    \includegraphics[width=0.9\textwidth]{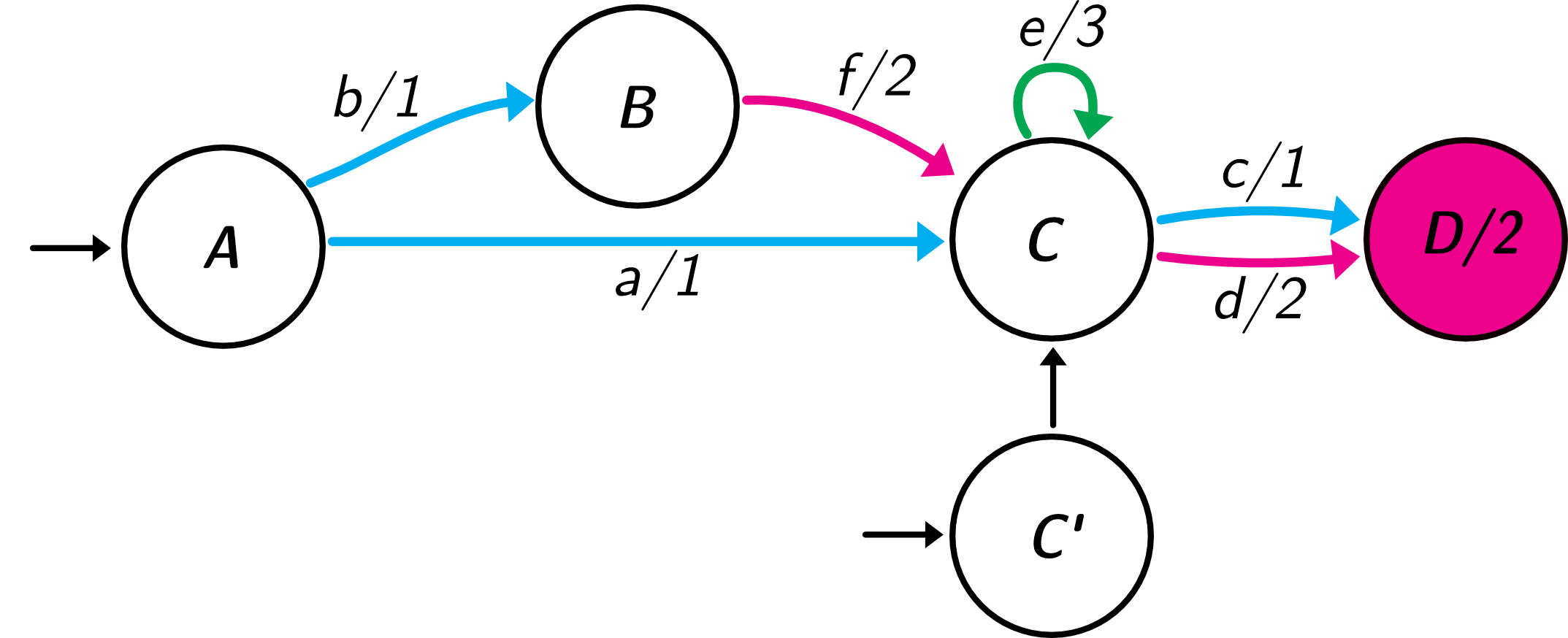}
    \caption{A call graph with several entry points. Node $C'$ is a sentinel
      vertex that forwards external calls to the internal vertex $C$. The call
      string $\mathit{ac}$ has been taken, immediately followed by call string
      $c$. The current context string is $c$.}\label{fig:library-2}
  \end{subfigure}
  \caption{Conceptual illustration of how the call history is maintained by
    coloring the call graph during runtime.}\label{fig:color}
\end{figure}

\begin{example}\label{ex:coloring-setup}
  Figure~\ref{fig:color-init} shows the initial coloring state of the call graph
  in Figure~\ref{fig:call-graph}.
  For instance, the vertex~$C$ has three incoming edges (with colors
  blue~($1$), magenta~($2$), and green~($3$)), while~$D$ has two (blue and
  magenta).
  Note that we can reuse a given color for several edges, as long as each vertex
  does not have two incoming edges with the same color. For instance, the edges
  labeled $a$ and $b$ are both blue. \qed
\end{example}

Algorithm~\ref{algo:gc-traverse} describes the update to the coloring when an
edge is traversed in the call graph.
\begin{algorithm}
  \caption{Traversing an edge.}\label{algo:gc-traverse}
  \textbf{Input} A call graph $G = (V,E,L,S,H)$, coloring functions $\ColorV$
  and $\ColorE$, and traversed edge $(v_1, v_2, l) \in E$.
  \\
  \textbf{Output} Modified coloring function $\ColorV'$. \\
  \begin{algorithmic}[1]
    \Procedure{TraverseEdge}{$G, \ColorV, \ColorE$, $(v_1,v_2,l)$}
    \State $\ColorV \gets \left( \ColorV \setminus \{v_2 \mapsto \ColorV(v_2)\} \right) \Union \{v_2 \mapsto \ColorE(l)\}$
    \Comment{Overwrite previous mapping of $v_2$ in
      $\ColorV$}\label{l:traverse-1}
    \State \Return $\ColorV$
    \EndProcedure
  \end{algorithmic}
\end{algorithm}
Line~\ref{l:traverse-1} updates the color of the destination vertex to the color
of the label of the edge being traversed.

\begin{example}\label{ex:coloring-traverse}
  Figure~\ref{fig:color-step-1} shows the state of the call graph after the call
  string $\mathit{ad}$ is taken, and Figure~\ref{fig:color-step-2} shows the
  state after the call string $\mathit{bfed}$ is taken. \qed
\end{example}

When the value of a hole is used during runtime of the program, we check the
current context by following the colors of the vertices and edges backwards in
the graph. The tracing stops when we reach the depth of the hole, when we reach
a white vertex, or when we detect a cycle.

\begin{example}\label{ex:coloring2}
  To determine the current context string in Figure~\ref{fig:color-step-1}, we
  first inspect the color of $D$ (magenta), which means that $d$ is the last
  label in the string. Next, we see that $C$ is blue, so $C$ was called by $A$,
  thus $a$ is the second to last label. The color of $A$ is white, so we stop
  the tracing. Thus, the context string is $ad$. \qed
\end{example}

\begin{example}\label{ex:coloring3}
  Similarly, in Figure~\ref{fig:color-step-2} we determine that the last two
  labels are $d$ and $e$. Since $C$ called itself via $e$, we have detected a
  cycle. Thus, the context string is $ed$. \qed
\end{example}

\subsubsection{Code Libraries}
  \label{sec:library}
  \sloppypar{%
    In contrast to a complete program, a separately compiled code library may
    have several entry points, namely, the publicly exposed functions. Further,
    each entry point may have incoming edges from internal calls within the
    library. This poses a problem for the coloring scheme described thus far,
    illustrated in the following example.
  \begin{example}\label{ex:library}
    Assume that vertex $C$ in Figure~\ref{fig:color-init} also is an entry
    point, along with vertex $A$. Then the set of context strings listed in
    Section~\ref{sec:eq-rel} is extended with $\mathit{c}$ and $\mathit{d}$. The reason for this is that a call directly to $C$ can result in the path $c$ or path $d$, without visiting any other edges.
    Further, assume that the call string $\mathit{ac}$ has been taken
    immediately before a call to $C$ is made and the call string $c$ is taken. Then the coloring state of
    the graph would be like in Figure~\ref{fig:color-step-1}, if we follow the
    coloring scheme described thus far. From this state, we cannot determine
    whether the current context string is $\mathit{ac}$ or $c$.
    \qed
  \end{example}

  A possible solution to this problem is shown in
  Figure~\ref{fig:library-2}. For each library node ($C$ in this
  example), we add an sentinel vertex (noted with a prime) which
  directly connects to the original entry point of the library. If a
  call is via an sentinel vertex, the next vertex is colored
  white. Hence, it is possible to distinguish between if the call is
  coming from the library entry point or from another vertex in the
  graph.

\subsection{Implementation of Graph Coloring}
\label{sec:impl-transformations}

\newcommand{\TransP}[0]{\Program_t}

The input to the program transformations is a program $\Program$ with holes, as
well as the path to the tune file. The output is a transformed program $\TransP$
that performs graph coloring, and where each base hole in $\Program{}$ has been
replaced by code that statically looks up a value depending on the current
context.
We discuss the program transformations in Section~\ref{sec:trans-static}, and
analyze the runtime overhead of the transformed program in
Section~\ref{sec:overhead}.

\subsubsection{Program Transformations}
\label{sec:trans-static}

During compile-time, we build a call graph $G$ as defined in
Section~\ref{sec:context-intuition}. A key idea is that we do need to maintain the call graph during runtime of the program; it is only used for analysis during the program
transformations.

In the transformed program, we introduce for
each vertex $v\in V'$, an integer reference $r_v$ whose initial value is $0$
(white).

The $\textsc{TraverseEdge}$ procedure in Algorithm~\ref{algo:gc-traverse} is
then implemented as a transformation. Immediately before a function call from $v_1$
to a function $v_2$, where $(v_1, v_2,l)\in E'$, we introduce an update of the
reference $r_{v_2}$ to the color (i.e., integer value) that the edge $(v_1,
v_2,l)$ is assigned to.

Finally, determining the current context string, as informally described in
Examples~\ref{ex:coloring2} and~\ref{ex:coloring3} also requires a program
transformation, which we call \emph{context expansion}. In context expansion, we
replace each base hole in the program with code that first looks up the current
context string, and thereafter looks up the value of the associated context
hole. Determining the current context string for a hole of depth $d$ requires
checking the values of at most $d$ integer references. For example, if the
following declaration of a base hole exists in vertex $D$ in
Figure~\ref{fig:call-graph}:
\begin{mcore-lines}
hole (Boolean {default = true, depth = 3})
\end{mcore-lines}
then we replace it by the following program code:
\begin{mcore-lines}
switch deref $r_D$
case 1 then
  switch deref $r_C$
  case 1 then <lookup $\mathit{ac}$>
  case 2 then <lookup $\mathit{bfc}$>
  case 3 then <lookup $\mathit{ec}$>
  end
case 2 then
  switch deref $r_C$
  case 1 then <lookup $\mathit{ad}$>
  case 2 then <lookup $\mathit{bfd}$>
  case 3 then <lookup $\mathit{ed}$>
  end
end
\end{mcore-lines}
where \mcoreinline{deref} reads the value of a reference, each $r_v$ is the
reference storing the color of function $v$, and each \mcoreinline{<lookup
  $\phantom{s}s$>} is code that looks up the value for the context hole
associated with the context string $s$.
In the final tuned program, each \mcoreinline{<lookup $\phantom{s}s$>} is simply
a static value: the value that has been tuned for context $s$ (tuned
compilation, see Section~\ref{sec:implementation}).
During tuning of the program, each \mcoreinline{<lookup $\phantom{s}s$>} is an
access into an array that stores the values of the context holes contiguously.
This array is read as input to the program via the tune file, so that the
program does not have to be re-compiled during tuning (see
Section~\ref{sec:implementation} for more details).
Note that a global hole (depth $0$) can be seen as having one context string,
namely the empty string, and thus does not need any \mcoreinline{switch}
statement.

\subsubsection{Adaption to Parallel Execution}
\label{sec:parallel}

In a parallel execution setting, there might be more than one active call string
during each given time in the program. We make an adaption to the program
transformation in order to handle a fixed number of threads $T$.
Instead of introducing \emph{one} reference per (relevant) function, we
introduce an array with $T$ number of references per (relevant) function. Each
thread $t$ is assigned an array index $t_i$. Each thread uses the references at
index $t_i$ only. In this way, we maintain up to $T$ active context strings
simultaneously.
If a thread pool is used, then the size of the thread pool needs to be
known at compile-time.
Otherwise, the transformation works as described in
Section~\ref{sec:trans-static}.

\subsubsection{Runtime Overhead in the Resulting Program}
\label{sec:overhead}

\newcommand{\ProgramDef}[0]{\Program_{\mathit{def}}}

As a baseline for runtime overhead, consider the original program $\Program{}$
where each base hole is replaced by its default value (default compilation, see
Section~\ref{sec:implementation}): we call this program $\ProgramDef$.
The overhead of the transformed program $\TransP$ compared to $\ProgramDef{}$
when the number of threads $T = 1$, includes initializing at most one integer
reference per function.
The program $\TransP$ performs at most one reference update per function call.
Moreover, $\TransP$ performs at most $d$ number of matches on references in
\mcoreinline{switch} statements each time it uses the value of a hole, where $d$
is the depth of the hole.
The underlying compiler can transform the \mcoreinline{switch} statements into
an indexed lookup table. This lookup table is compact by construction, as we
use contiguous integers as values representing colors.
When $T > 2$, then $\TransP$ introduces at most one array of references per
function, and each reference update and reference read includes an indexing into
an array.

\section{Static Dependency Analysis}\label{sec:dependency-analysis}

This section discusses static dependency analysis. The goal is to detect holes
that can be tuned independently of each other. This information is later used
during tuning in order to reduce the search space.
Section~\ref{sec:dep-motivation} motivates the need of dependency analysis and
provides intuition, Section~\ref{sec:dep-definitions} makes necessary
definitions that are used in Section~\ref{sec:dep-analysis}, which describes the
details of the dependency analysis. Finally,
Section~\ref{sec:dep-instrumentation} describes how the program is instrumented
given the result of the dependency analysis.

\subsection{Motivation and Running Example}\label{sec:dep-motivation}

Consider the following $k$-nearest neighbor ($k$-NN) classifier, which will be
our running example in this section:
\begin{lstlisting}[language=MCore,firstline=38,lastline=51]
let knnClassify = lam k: Int. lam data: [([Int],Label)]. lam query: [Int].
  -- Step 1: compute the distance to each point in the data set
  let dists: [(Int,Label)] = map (lam d: ([Int],Label).   --*\label{l:knn-dists-start}*--
      (euclideanDistance query d.0, d.1)
    ) data  --*\label{l:knn-dists-end}*--
  in
  -- Step 2: sort the distances in ascending order
  let sortedDists: [(Int,Label)] = sort (
      lam d1: (Int,Label). lam d2: (Int,Label). subi d1.0 d2.0
    ) dists
  in
  -- Step 3: return the most common label among the k nearest neighbors
  let kNearest: [(Int,Label)] = subsequence sortedDists 0 k in  --*\label{l:knn-subsequence}*--
  mostCommonLabel kNearest
\end{lstlisting}
The classifier takes three arguments:
the parameter \mcoreinline{k} of the algorithm;
the \mcoreinline{data} set, which is a sequence of tuples $(d,l)$, where $d$ is
a data point (representing an integer vector) and $l$ is the class label; and
the \mcoreinline{query} data point, whose class label we want to decide.

The algorithm has three steps.
In the first step, we compute the pairwise distances between the query and each
point in the data set, in this example using Euclidean distance.
In step two, we sort the pairwise distances. The first argument to the function
\mcoreinline{sort} is the comparison function, which in this case computes the
difference between two distances.
In the last step, we extract the $k$ nearest neighbors by taking the $k$ first
elements in the sorted sequence \mcoreinline{sortedDists}. Finally, we assume
that the \mcoreinline{mostCommonLabel} function returns the most frequent label
in a sequence of labels, so that the query point is classified to the most
common class among its neighbors.

\newcommand{\Hseq}[0]{h_{\text{seq}}}
\newcommand{\Hmap}[0]{h_{\text{map}}}
\newcommand{\Hsort}[0]{h_{\text{sort}}}

Now, assume that the $k$-NN classifier implicitly uses three holes.
The first hole, $\Hseq$, is for deciding the underlying data structure for the
sequences. In the Miking core language, a sequence can either be represented by
a cons list, or a Rope~\cite{ropes}. We can use a Boolean hole to choose the
representation when creating the sequence, by either calling the function
\mcoreinline{createList} or \mcoreinline{createRope}.
The second hole, $\Hmap$, chooses between sequential or parallel code in the
\mcoreinline{map} function, see Example~\ref{example-map2}.
Finally, the third hole, $\Hsort$, chooses between two sorting algorithms
depending on an unknown threshold value, see Example~\ref{ex:global}.

When tuning the classifier, these three choices need to be taken into
consideration. If the program is seen as a black box, then an auto tuner needs
to consider the \emph{combination} of each of these choices.
In this small example, we quickly see that while some choices are indeed
necessary to explore in combination, others can be explored in isolation.
Choices that must be explored together, called \emph{dependent} choices, are for
instance:
(i) the underlying sequence representation and the \mcoreinline{map} function
($\Hseq$ and $\Hmap$); and (ii) the underlying sequence representation and the
\mcoreinline{sort} function ($\Hseq$ and $\Hsort$).
In both cases, this is because the sequence representation affects the execution
time of the operations performed on the sequence in the respective function.
On the other hand, the \mcoreinline{sort} function and the \mcoreinline{map}
function do \emph{not} need to be explored in combination with each other: the
holes $\Hmap$ and $\Hsort$ are \emph{independent} of each other. Regardless of
what choice is made in the \mcoreinline{map} function (sequential or parallel),
the result of the function is the same, which means that the \mcoreinline{sort}
function should be unaffected.\footnote{We say \emph{should} here as cache
  effects from \mcoreinline{map} may still affect the execution time of
  \mcoreinline{sort}.}

With knowledge about independencies, an auto tuner can use the tuning time in a
more intelligent way, as it does not need to waste time exploring unnecessary
combinations of holes.
The remainder of this section describes how we can automatically detect
(in-)dependencies such as the examples discussed here, using static analysis.

\subsection{Definitions}\label{sec:dep-definitions}

Before discussing the details of the dependency analysis, we need to define the
entities that constitute dependency: measuring points and dependency graphs.

\subsubsection{Measuring Points}\label{sec:dep-measuring-points}

Intuitively, two holes are independent if they affect the execution time of
disjoint parts of the program. That is, we want to find the set of
subexpressions of the program whose execution times are affected by a given
hole. There are often many such subexpressions. For instance, the complete
\mcoreinline{knnClassify} in Section~\ref{sec:dep-motivation} is a subexpression
whose execution time depends on three holes: $\Hseq$, $\Hmap$, and $\Hsort$.
Moreover, the subexpression on
Lines~\ref{l:knn-dists-start}--\ref{l:knn-dists-end} (the computation of
\mcoreinline{dists}) in \mcoreinline{knnClassify} depends on $\Hseq$ and
$\Hmap$.
So how do we choose which subexpressions that are relevant in the dependency
analysis?

Clearly, it is not useful to consider too large subexpressions of the program.
This is because two holes $h_1$ and $h_2$ may affect a large subexpression $e$,
even though they in reality only affect smaller, disjoint subexpressions $e_1$
and $e_2$, respectively, where $e_1$ and $e_2$ are subexpressions of $e$.
Therefore, we want to find small subexpressions whose execution time depends on
a given hole. We exemplify the type of expressions we are interested in for
\mcoreinline{knnClassify} in Example~\ref{ex:measuring-points}, before going
into details.

\begin{example}\label{ex:measuring-points}
  Assume that the \mcoreinline{map} function is given by
  Example~\ref{example-map2}, and the \mcoreinline{sort} function is given by
  Example~\ref{ex:global}.
  A small subexpression affected by~$\Hmap$ is Line~\ref{l:par-ite} in
  \mcoreinline{map} (the if-then-else expression), because which branch is taken
  is decided by the hole \mcoreinline{par}.
  Similarly, the if-then-else expression on
  Lines~\ref{l:sort-ite-1}--\ref{l:sort-ite-2} in \mcoreinline{sort} is a
  small subexpression affected by~$\Hsort$.
  These two subexpressions are also affected by $\Hseq$, because the execution
  times of the branches depend on the underlying sequence representation.
  Furthermore, Line~\ref{l:knn-subsequence} in \mcoreinline{knnClassify} is a
  minimal subexpression affected by $\Hseq$, because the execution time of
  \mcoreinline{subsequence} also depends on the underlying sequence
  representation.
  \qed{}
\end{example}

We call these small subexpression whose execution time depends on at least one
hole a \emph{measuring point}. The rationale of the name is that we measure the
execution time of these subexpressions by using instrumentation (see
Section~\ref{sec:dep-instrumentation}).
The Miking language, being a core language, consists of relatively few language
constructs; any higher-order language implemented on top of Miking will compile
down to this set of expressions. The type of expressions that construct
measuring points are either:
\begin{enumerate}
\item a match statement (including if-then-else expressions); or
\item a call to a function \mcoreinline{f x}, where \mcoreinline{f} is either a
  built-in function (such as \mcoreinline{subsequence}), or user-defined.
\end{enumerate}
Section~\ref{sec:dep-analysis} clarifies under which circumstances these
expressions are measuring points. Other types of expressions in Miking, such as
lambda expressions, constants, records, and sequence literals, are not relevant
for measuring execution time.

\subsubsection{Dependency Graph}\label{sec:dep-dependency-graph}

\begin{wrapfigure}{r}{0.4\linewidth}
    \centering
      % trim={<left> <lower> <right> <upper>}
    \includegraphics[trim={0 2cm 0
      0},clip,width=0.16\textwidth]{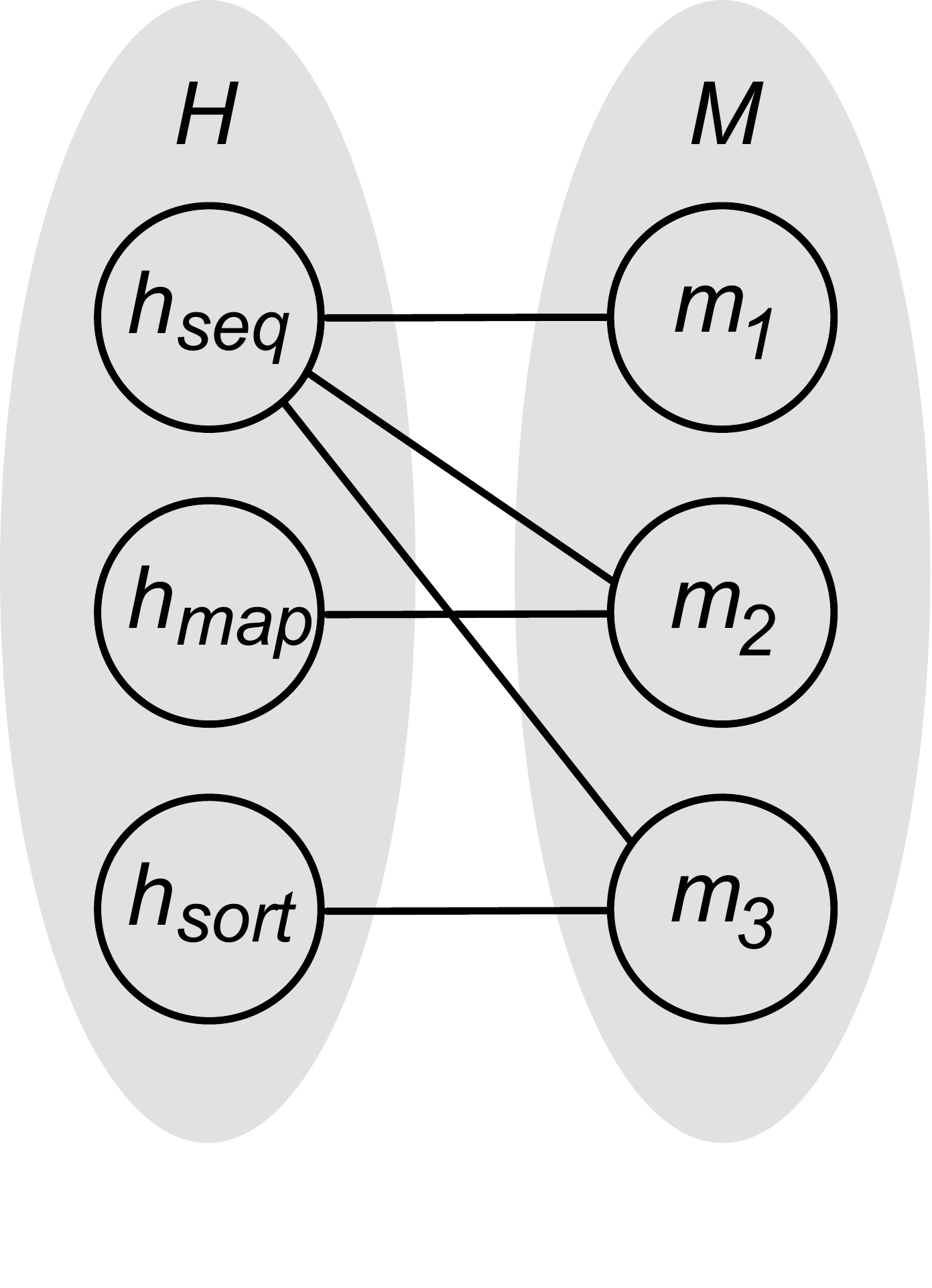}
      \caption{Dependency graph for the \mcoreinline{knnClassify} function in
        Section~\ref{sec:dep-motivation}. The set $H$ is the set of holes (the
        choice of sequence representation, choice of sequential or parallel map,
        and the choice of sorting function, respectively). The set $M$ is the
        set of measuring points, where $m_1$ is Line~\ref{l:knn-subsequence} of
        \mcoreinline{knnClassify}, point $m_2$ is Line~\ref{l:par-ite} of
        \mcoreinline{map}, and $m_3$ is
        Lines~\ref{l:sort-ite-1}--\ref{l:sort-ite-2} of
        \mcoreinline{sort}.}\label{fig:knn-dep-graph}
  \end{wrapfigure}

We now define dependency graphs.
Given a program $\Program{}$ with a set of context holes $\HolesSet{}$, and a
set of measuring points $\MeasSet{}$, its \emph{dependency graph} is a bipartite
graph $(\HolesSet{},\MeasSet{},E)$. There is an edge $(h,m) \in E$, $h\in H$,
$m\in M$, iff the hole $h$ affects the execution time of $m$.

\begin{example}
  A (partial) dependency graph of \mcoreinline{knnClassify} is given by
  Figure~\ref{fig:knn-dep-graph}. It is partial because there are more measuring
  points in the functions \mcoreinline{mostCommonLabel} and \mcoreinline{pmap}
  (because they contain sequence operations, whose execution times depend on
  $\Hseq$), but these functions are omitted for brevity.
  \qed{}
\end{example}

The dependency graph encodes the dependencies among the program holes. For
instance, in Figure~\ref{fig:knn-dep-graph}, we see that $\Hseq$ and $\Hmap$ are
dependent, because they both affect $m_2$, while $\Hmap$ and $\Hsort$ are
independent, because they have no measuring point in common.

\subsection{Dependency Analysis}\label{sec:dep-analysis}

The goal of the dependency analysis is to compute the dependency graph of a
given program.

\subsubsection{0-CFA analysis}\label{sec:cfa-analysis}

The backbone of the dependency analysis is $0$-CFA
analysis~\cite{principles-of-program-analysis}. We extend standard $0$-CFA
analysis, which tracks data-flow of functions, to additionally track data-flow
of holes.
The result is that we compute for each subexpression in the program the set of
holes that affect the \emph{value} of the subexpression.

Standard data-flow rules apply.
The first two columns of Table~\ref{tab:dep-data} shows the result of the
data-flow analysis for a few example subexpressions from
\mcoreinline{knnClassify}.
We denote the data dependency of a subexpression by the set of holes whose value
the subexpression depends on.
In the first row, the variable \mcoreinline{query} depends on $\Hseq$ because
the variable refers to a sequence whose representation is decided by $\Hseq$.
Second, the if-then-else expression from the \mcoreinline{map} function depends
on both $\Hseq$ and $\Hmap$, because the condition of the if-then-else depends
on $\Hmap$, and the result of the subexpression is again a sequence dependent on
$\Hseq$.
In the third row, the variable \mcoreinline{dists} also depends on both $\Hseq$
and $\Hmap$, because the variable refers to the result of the \mcoreinline{map}
function.
In the fourth row, Lines~\ref{l:sort-ite-1}--\ref{l:sort-ite-2} in
Example~\ref{ex:global} depends on all three holes. It depends on $\Hsort$
because the condition of the if-then-else depends on $\Hsort$. It depends on
$\Hseq$ and $\Hmap$ because the branches of the if-then-else manipulate the
sequence referred to by \mcoreinline{dists}.
Finally, the call to \mcoreinline{subsequence} also depends on all holes,
because the built-in function \mcoreinline{subsequence} returns a sequence that
will have the same data dependency as its input sequence,
\mcoreinline{sortedDists}.

\begin{table}
  \centering
  \caption{Result of the dependency analysis for a subset of the subexpressions
    in \mcoreinline{knnClassify}, using Examples~\ref{ex:global} and
    \ref{example-map2} as implementations of \mcoreinline{sort} and
    \mcoreinline{map}, respectively.
    Columns~$2$--$3$ and $4$--$5$ show the data dependency and the execution
    time dependency of each subexpression, for a program without and with
    annotations, respectively.
    We see that without the annotation, the analysis concludes that all three
    holes are execution time dependent, while the version with annotations
    results in the dependency graph in Figure~\ref{fig:knn-dep-graph}.
  }\label{tab:dep-data}
  % \begin{tabular}{c|l|l|l|l}
  \begin{tabular}{ccccc}
    & \multicolumn{2}{c}{Without annotations} & \multicolumn{2}{c}{With annotations} \\
    \cmidrule(lr){2-3}\cmidrule(lr){4-5}
    Subexpression & Data dep. & Exe. dep. & Data dep. & Exe. dep. \\
    \midrule
    \mcoreinline{query} & $\{\Hseq\}$ & $\varnothing$ & $\{\Hseq\}$ & $\varnothing$ \\
    \mcoreinline{if par then ... else ...} & $\{\Hseq,\Hmap\}$ & $\{\Hseq,\Hmap\}$ & $\{\Hseq\}$ & $\{\Hseq,\Hmap\}$ \\
    \mcoreinline{dists} & $\{\Hseq,\Hmap\}$ & $\varnothing$ & $\{\Hseq\}$ & $\varnothing$ \\
    Lines~\ref{l:sort-ite-1}--\ref{l:sort-ite-2} in Example~\ref{ex:global} & $\{\Hseq,\Hmap,\Hsort\}$ & $\{\Hseq,\Hmap,\Hsort\}$ & $\{\Hseq\}$ & $\{\Hseq,\Hsort\}$ \\
    \mcoreinline{subsequence sortedDists 0 k} & $\{\Hseq,\Hmap,\Hsort\}$ & $\{\Hseq,\Hmap,\Hsort\}$ & $\{\Hseq\}$ & $\{\Hseq\}$ \\
    %\mcoreinline{sortedDists} & $\{\Hseq,\Hmap,\Hsort\}$ & $\varnothing$ & $\{\Hseq\}$ & $\varnothing$ \\
  \end{tabular}
\end{table}

Recall that we are interested in subexpressions whose \emph{execution time}
(\emph{not} value) depends on holes: these are the measuring points of the
program. Luckily, we can incorporate the analysis of measuring points into the
$0$-CFA, by using the data-flow information of the holes.
Besides data dependency, we introduce another kind of dependency:
\emph{execution time dependency}.
A subexpression with a non-empty execution time dependency is a measuring point.
There are two kinds of expressions that may give rise to execution time
dependency, i.e., measuring points: match expressions, and calls to functions.

\paragraph{Match Expressions.}
Given a match expression \mcoreinline{e} on the form \mcoreinline{match e1 with
  pat then e2 else e3}, the following two rules apply:
%\begin{enumerate}
(1) If \mcoreinline{e1} is \emph{data}-dependent on a hole $h$, then
  \mcoreinline{e} is \emph{execution time}-dependent on $h$; and
(2) If \mcoreinline{e2} executes (directly or via a function call) another
  subexpression that is \emph{execution time}-dependent on a hole $h$, then
  \mcoreinline{e} is also execution time-dependent on $h$, and the same applies
  for \mcoreinline{e3}.
%\end{enumerate}
%
The justification of rule 1 is that if the decision of which branch to take
depends on a hole, then the execution time of the match expression depends on
the hole. The justification of rule 2 is that the execution time of the whole
subexpression \mcoreinline{e} should include any execution time dependencies of
the individual branches.

The third column of Table~\ref{tab:dep-data} shows execution time dependencies
of some subexpressions in \mcoreinline{knnClassify}. Rows 2 and 4 are match
expressions.
Note that in the Miking core language, an if-then-else expression is syntactic
sugar for a match expression where \mcoreinline{pat} is \mcoreinline{true}.
The conditions of the match expressions depend on $\Hmap$ and $\Hsort$,
respectively, thus these holes are included in the execution time dependencies.
The branches of each subexpression perform sequence operations, which will be
measuring points dependent on $\Hseq$. Thus, the execution time of the match
expressions also depends on $\Hseq$.
The dependency on row 4 also includes $\Hmap$, because the input sequence,
\mcoreinline{dists}, has a data dependency on $\Hmap$.

\paragraph{Function calls.}

If the expression \mcoreinline{e} is the result of an application of a
\emph{built-in} function, then custom rules apply for each built-in. For
instance, for the expression \mcoreinline{subsequence s i j}, if the sequence
\mcoreinline{s} is \emph{data-dependent} on a hole $h$, then \mcoreinline{e} is
\emph{execution time}-dependent on $h$.
The \mcoreinline{subsequence} expression in the last row of
Table~\ref{tab:dep-data} has the same execution time dependency as its data
dependency, by following this rule.

In addition, for all function calls \mcoreinline{e} on the form \mcoreinline{e1
  e2}, if \mcoreinline{e1} is \emph{data}-dependent on a hole $h$, then
\mcoreinline{e} is \emph{execution time}-dependent on $h$. As a simple example,
the function call \mcoreinline{(if h then f else g) x} is a measuring point,
given that \mcoreinline{h} is data-dependent on some hole. In other words, since
the left hand side of the application is determined by a hole, the execution
time of the function call depends on a hole.

Note that the function call on
Lines~\ref{l:knn-dists-start}--\ref{l:knn-dists-end} in
\mcoreinline{knnClassify} does \emph{not} constitute a measuring point, even
though its execution time depends on $\Hseq$ and $\Hmap$. The reason is that the
function \mcoreinline{map} itself is not data-dependent on a hole. The relevant
execution times of the \mcoreinline{map} functions are already captured by
measuring points within the \mcoreinline{map} function.

\subsubsection{Call Graph Analysis}\label{sec:call-graph-after-cfa}

The $0$-CFA analysis finds the set of measuring points of the program and
attaches an initial set of dependencies to each measuring point. Some
dependencies, however, are not captured in the $0$-CFA analysis. Namely, if a
measuring point $m_1$ executes \emph{another} measuring point $m_2$, then the
holes that affect $m_2$ also affect $m_1$.
For instance, the expression \mcoreinline{if par then pmap f s else smap f s}
executes any measuring points within the \mcoreinline{pmap} and
\mcoreinline{smap} function.
We perform another analysis step that analyzes the call graph of the program, in
order to find the set of measuring points that each measuring point executes.
This analysis step does not introduce any new measuring points, but it adds
more dependencies (edges in the dependency graph).

\subsubsection{False Positives and Annotations}\label{sec:annotations}

As we have seen so far, the dependencies (both for data and execution time) on
some subexpressions in Table~\ref{tab:dep-data} are unnecessarily large. For
instance, the \emph{value} of the subexpression on row 2 should intuitively
\emph{not} depend on $\Hmap$. After all, whether the map is performed in
parallel or sequentially does not affect the final result.
In other words, the data dependency on $\Hmap$ on row 2 is a false positive.

The result of false positives on data dependencies is that some execution time
dependencies may also be unnecessarily large. As we see in
Table~\ref{tab:dep-data}, the false positive on $\Hmap$ on row 2 propagates to
the data dependency of row 3 (\mcoreinline{dists}), which in turn affects the
\emph{execution time} dependencies of rows 4 and 5.

While it is in general hard for a compiler to detect, for instance, that
parallel and sequential code gives the same end result, or that two sorting
functions are equivalent, this information is typically obvious for a
programmer.

Therefore, we introduce the option to add \emph{annotations} to a program to
reduce the number of false-positive dependencies. The annotation states the set
of variables that a match expression is independent of, and is added directly
after a match expression using the keyword \mcoreinline{independent}.

For instance, replacing Line~\ref{l:par-ite} in Example~\ref{example-map2} with
\mcoreinline{independent (if par then pmap f s else smap f s) par}
states that the value of the match expression is independent of the variable
\mcoreinline{par}. More variables can be included in the set by nesting several
\mcoreinline{independent} annotations, e.g. \mcoreinline{independent (independent
  <e> x) y}.

By incorporating this information in the analysis, the data dependency on the
independent set is ignored for the match expression.
Columns $4$--$5$ in Table~\ref{tab:dep-data} show the result of the analysis
given that the match expressions on rows 2 and 4 have been annotated to be
independent of the variables \mcoreinline{par} and \mcoreinline{threshold},
respectively. We see that the execution time dependencies now match the
dependency graph in Figure~\ref{fig:knn-dep-graph}. Row 2 in
Table~\ref{tab:dep-data} corresponds to $m_2$, row 4 corresponds to $m_3$, and
row 5 corresponds to $m_1$.

\subsubsection{Context-Sensitive Measuring Points}\label{sec:dep-context}

A property of $0$-CFA is that it does not include context information for the
data-flow, unlike $k$-CFA for $k>0$. While we are limited to $0$-CFA for
efficiency reasons, it is necessary to consider the contexts of
context-sensitive holes.
Therefore, we consider the context strings (see
Section~\ref{sec:transformations}) during the dependency analysis.

As an example, consider the \mcoreinline{map} function in
Example~\ref{example-map2}. Assume that it is called from two locations, so that
there are two possible call strings for the hole \mcoreinline{par}; $s_1$ and
$s_2$. During analysis of the measuring point on Line~\ref{l:par-ite}, we
conclude that the execution time depends \emph{either} on the context hole
associated with $s_1$, \emph{or} the one associated with $s_2$, but not both.
This is taken into account during instrumentation of the program, see
Section~\ref{sec:dep-instrumentation}.

\subsection{Instrumentation}\label{sec:dep-instrumentation}

The instrumentation is a program transformation step, where the input is the
program $p$ and the dependency graph $G=(\HolesSet{},\MeasSet{},E)$. The output
is an instrumented program $p_I$ that collects execution time information for
each measuring point.
Section~\ref{sec:instrumentation-challenges} introduces three challenges when
designing the instrumentation. In Section~\ref{sec:instrumentation-overview}, we
present the proposed design and clarify how the design addresses the identified
challenges.

\subsubsection{Challenges}\label{sec:instrumentation-challenges}
% Problems with naive solution
Assume that we wish to instrument the measuring point in row 2 in
Table~\ref{tab:dep-data} on page~\pageref{tab:dep-data}: \mcoreinline{if par
  then pmap f s else smap f s}. A naive approach is to save the current time
before and after the expression has been executed, and then record the elapsed
time after the expression has been executed.

However, there are a number of problems with this simple solution.
% Nested
First, the measuring point can execute another measuring point. In this specific
case, it executes any measuring points within the \mcoreinline{pmap} or
\mcoreinline{smap} functions. If we do not keep track of whether a measuring
point executes within another one, we will count some execution times several
times, which gives an inaccurate total execution time of the program.
% Tail-call optimization
Second, this simple instrumentation approach does not allow for tail-call
optimizations. The reason is that after the transformation of the program, some
operations are performed \emph{after} the execution of the measuring point. The
result is that a recursive call within the measuring point will no longer be in
tail position.
% Context-sensitivity
The third challenge has to do with context-sensitivity. For instance, assume
that \mcoreinline{map} in Example~\ref{example-map2} is called from two
locations. The instrumented code must then consider these two calling contexts
when recording the execution of the measuring point.

\subsubsection{Solution}\label{sec:instrumentation-overview}

The instrumentation introduces a number of global variables and functions in the
program, maintaining the current execution state via a lock mechanism. Moreover,
every measuring point is uniquely identified by an integer. In particular:
\begin{itemize}
\item
  The variable \mcoreinline{lock} stores the identifier of the measuring
  point that is currently running, where the initial value $0$ means that no
  measuring point is running.
  Note that we do not mean a lock for parallel execution; we can still execute
  and measure parallel execution of code.
\item
  The array \mcoreinline{s} of length $|M|$, where \mcoreinline{s}$[i]$
  stores the latest seen start time of the $i$th measuring point.
\item
  The array \mcoreinline{log} of length $|M|$, where \mcoreinline{log}$[i]$
  stores a tuple $(T,n)$ where $T$ is the accumulated execution time, and $n$ is
  the number of runs, of the $i$th measuring point.
\item
  The function \mcoreinline{acquireLock} takes an integer $i>0$ (a unique
  identifier of a measuring point) as argument, and is called upon entry of a
  measuring point. If the \mcoreinline{lock} equals $0$, then the function sets
  \mcoreinline{lock} to $i$, and writes the current time to
  $\mcoreinline{s}[i]$. Otherwise, that is, if the \mcoreinline{lock} is already
  taken, the function does nothing.
\item
  The function \mcoreinline{releaseLock} also takes an integer identifier
  $i>0$ as argument, and is called when a measuring point exits. If the
  \mcoreinline{lock} equals $i$, then the function sets \mcoreinline{lock} to
  $0$, and adds the elapsed execution time to the global \mcoreinline{log}. If
  the \mcoreinline{lock} is taken by some other measuring point, the function
  does nothing.
\end{itemize}
After the instrumented program is executed, the array \mcoreinline{log} stores
the accumulated execution time and the number of runs for each measuring point.

As a result, the measuring point in row 2 in
Table~\ref{tab:dep-data} is replaced by the following lines of code:
\begin{mcore-lines}
acquireLock $\mathit{i}$; --*\label{l:instr-1}*--
let v = if par then pmap f s else smap f s in  --*\label{l:instr-2}*--
releaseLock $\mathit{i}$; --*\label{l:instr-3}*--
v
\end{mcore-lines}
The \mcoreinline{lock} design addresses the first of the identified challenges:
to keeping track of whether a measuring point is executed within another one.
Because only one measuring point can possess the lock at any given moment, we do
not record the execution if one executes within another. Thus, the sum of the execution times in the \mcoreinline{log} array
never exceeds the total execution time of the program.
 
We now consider the second challenge: allowing for tail-call optimization.
For a measuring point with a recursive call \mcoreinline{f x} in tail position,
for instance \mcoreinline{if <$\mathit{cond}$> then <$\mathit{base case}$> else f x},
the call to \mcoreinline{releaseLock} is placed in the base case only, so that
the recursive call remains in tail position:

\begin{mcore-lines}
acquireLock $\mathit{i}$;
if <$\mathit{cond}$> then
  let v = <$\mathit{basecase}$> in
  releaseLock $\mathit{i}$;
  v
else f x
\end{mcore-lines}
There can be more than one call to \mcoreinline{releaseLock} in each base case,
because of measuring points in mutually recursive functions.
The instrumentation analyzes the (mutually) recursive functions within the
program and inserts the necessary calls to \mcoreinline{releaseLock} in the base
cases of these functions.

The third challenge, dependency on context-sensitive holes, is addressed
similarly as in Section~\ref{sec:trans-static}. Consider again the measuring
point \mcoreinline{if par then pmap f s else smap f s in} within the
\mcoreinline{map} function, and assume that there are two possible calls to
\mcoreinline{map}. The measuring point is assigned a different identifier
depending on which of these contexts is active. The identifier is found by a
\mcoreinline{switch} expression of depth 1, reading the current color of the
\mcoreinline{map} function.
If there is only one call to \mcoreinline{map}, then the identifier is simply an
integer, statically inserted into the program.

\section{Dependency-Aware Tuning}\label{sec:dep-tuning}

In contrast to standard program tuning, dependency-aware tuning takes the
dependency graph into account to reduce the search space of the problem.
This section describes how to explore this reduced search space, and how to find
the expected best configuration given a set of observations.

\subsection{Reducing the Search Space Size}\label{sec:reduce-search-space}

\begin{figure}
    \begin{subfigure}[t]{.30\textwidth}
      \centering
      \includegraphics[width=0.5\textwidth]{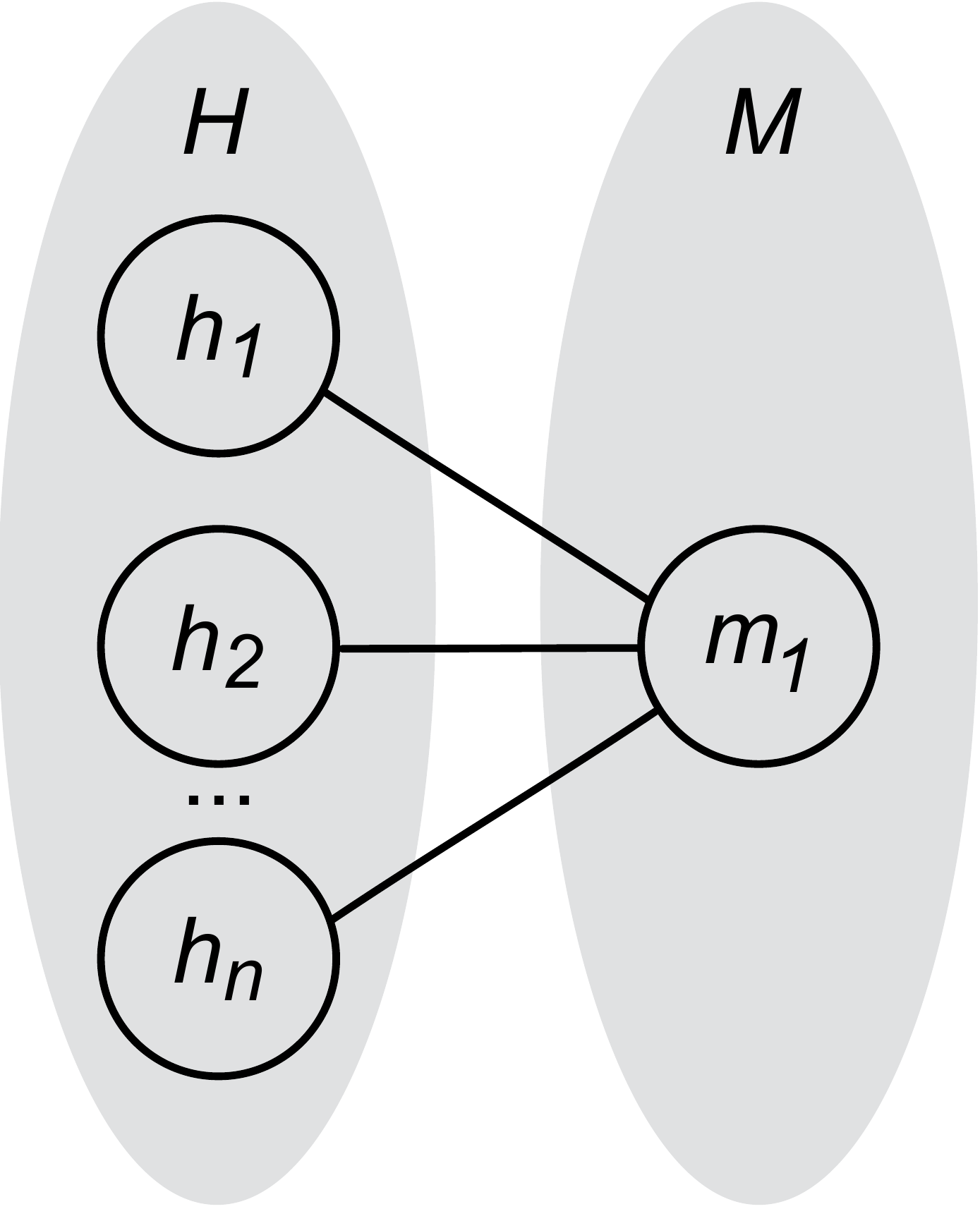}
      \caption{Dependency graph of a program with fully dependent
        holes. All $2^n$ possible combinations need to be taken into
        consideration when tuning.}\label{fig:bipartite-dependent}
    \end{subfigure}\hfill
    \begin{subfigure}[t]{.30\textwidth}
      \centering
      \includegraphics[width=0.5\textwidth]{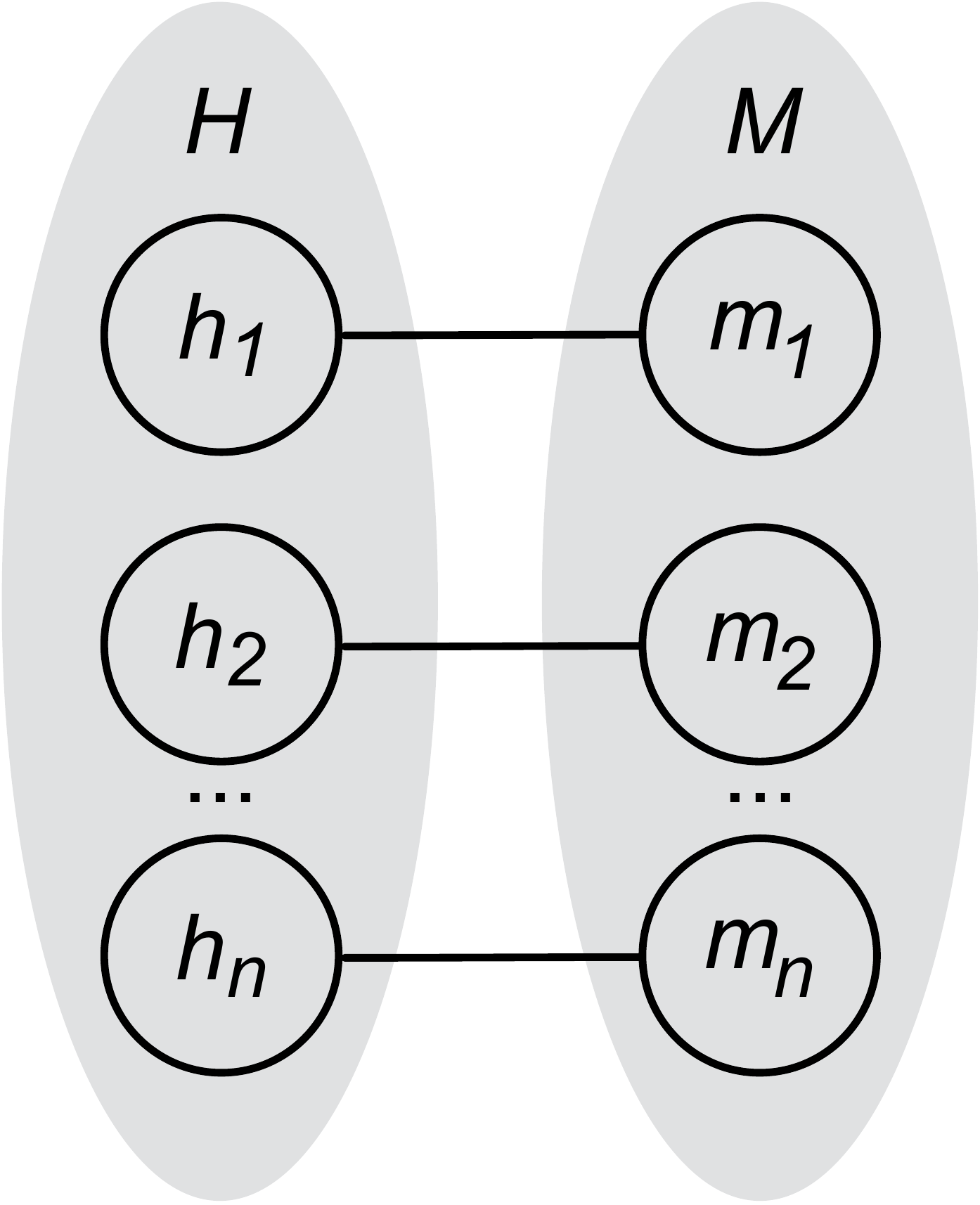}
      \caption{Dependency graph of a program with fully \emph{in}dependent
        holes. It is enough with $2$ program runs to exhaust the search space if
        fine-grained instrumentation of each measuring point is used.
        % Using
        % end-to-end execution time measurement requires $n+1$ program runs to
        % exhaust the search space.
      }\label{fig:bipartite-independent}
    \end{subfigure}\hfill
    \begin{subfigure}[t]{.30\textwidth}
      \centering
      \includegraphics[width=0.5\textwidth]{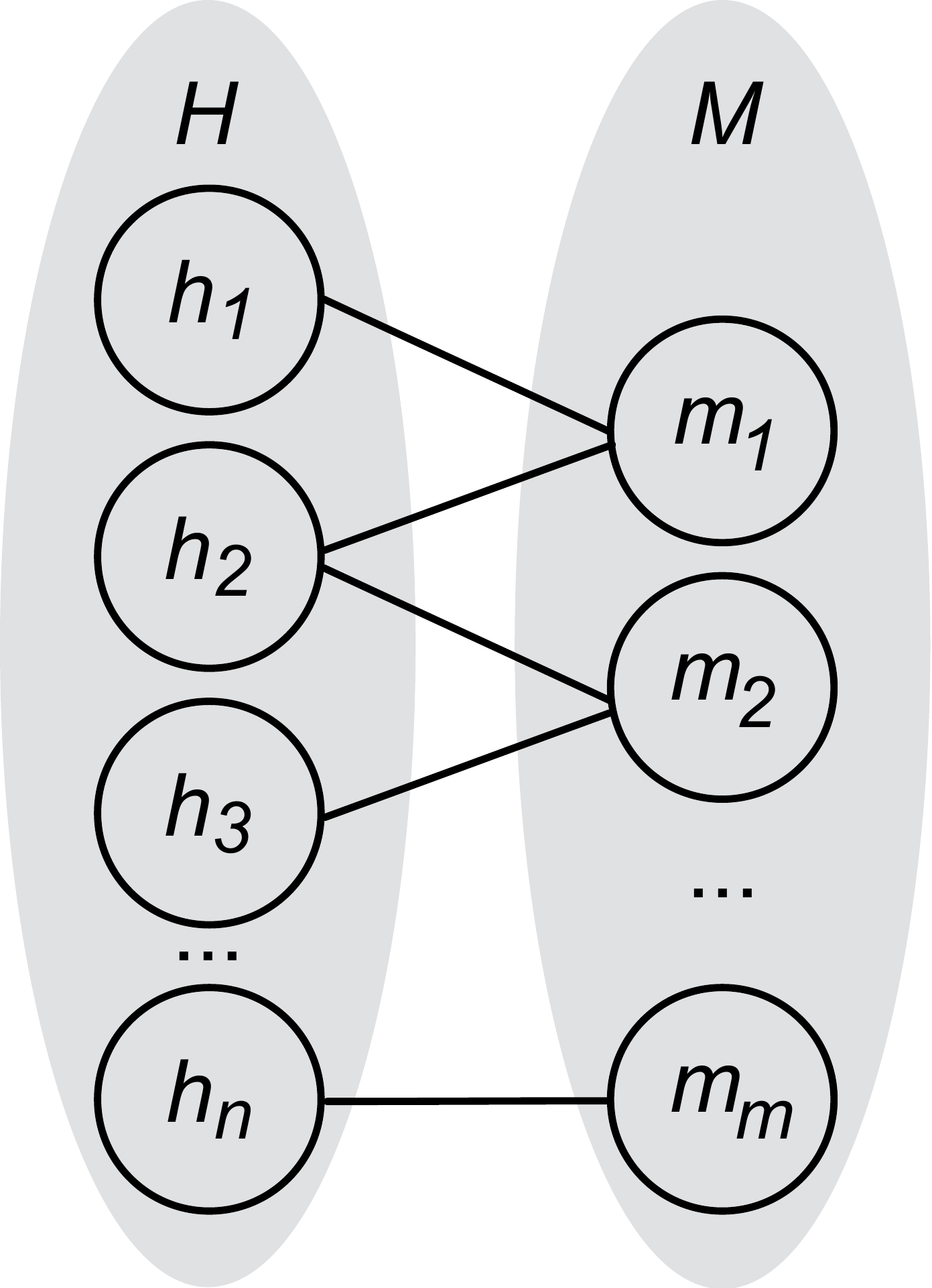}
      \caption{Dependency graph that is neither fully dependent nor fully
        independent. The number of required program runs with instrumentation is
        $2^r$, where $r$ is the maximum number of holes affecting any given
        measuring point. }\label{fig:bipartite-partial}
    \end{subfigure}
    \caption{Possible dependency graphs for a program with $n$ holes of Boolean
      type. In all cases, there are $2^n$ number of possible configurations.
      However, the more independence, the fewer of these configurations need to
      be considered when tuning.}\label{fig:bipartite}
  \end{figure}

In standard program tuning (without dependency analysis), each hole needs to be
tuned in combination with every other hole, which means that the number of
configurations to consider grows exponentially with the number of holes.

\begin{example}\label{ex:dep-fully-dep}
  Consider a program with $n$ Boolean holes, which has a search space of size
  $2^n$.
  With no dependency analysis, we may view the program as consisting of one
  measuring point, affected by all the holes in the program. This corresponds to
  the dependency graph in Figure~\ref{fig:bipartite-dependent}. If exhaustive
  search is used, then $2^n$ program runs are required to find the optimal
  configuration.
  \qed{}
\end{example}

Dependency analysis finds the fraction of the total number of configurations
that are relevant to evaluate during tuning, as illustrated in
Examples~\ref{ex:dep-fully-indep} and~\ref{ex:dep-partial}.

\begin{example}\label{ex:dep-fully-indep}
  Consider the program from Example~\ref{ex:dep-fully-dep}.
  Figure~\ref{fig:bipartite-independent} shows a dependency graph where all the
  holes are completely independent, so that each measuring point is affected by
  exactly one hole.
  If instrumentation is used, we collect the execution time for each measuring
  point in isolation. In this case, it is enough to run $2$ configurations to
  exhaust the search space. For example, we can run one configuration where all
  holes are set to \mcoreinline{true}, and one where they are set to
  \mcoreinline{false}. After this, the optimal configuration is found by
  considering the results for each hole in isolation and determining whether its
  value should be \mcoreinline{true} or \mcoreinline{false}.
  
  If end-to-end time measurement is used for the dependency graph in
  Figure~\ref{fig:bipartite-independent}, then $n+1$ program runs is required.
  For example, one run where all holes are set to \mcoreinline{false}, followed
  by $n$ runs where each hole at a time is set to \mcoreinline{true}, while
  keeping the remaining holes fixed.
  \qed{}
\end{example}  

\begin{example}\label{ex:dep-partial}
  Again considering the program from Example~\ref{ex:dep-fully-dep},
  Figure~\ref{fig:bipartite-partial} shows a scenario where the holes are
  neither fully dependent nor fully independent. Assume that $n=4$ and $m=3$, so
  that the dependency graph contains only the holes and measuring points that
  are visible (without the ``$\ldots$'' parts). There are at most $2$ holes that
  affect any given measuring point, and each hole has $2$ possible values.
  Therefore it is enough to consider $2\cdot 2=4$ configurations. For example,
  we may consider the ones listed in Table~\ref{tab:configurations}, though this
  table is not unique.
  Note that the table contains all combinations of values for $\{h1, h2\}$, for
  $\{h2, h3\}$, and for $\{h4\}$, respectively. However, some combinations of
  $\{h1,h3\}$ are missing, because $h1$ and $h3$ do not have any measuring point
  in common.
  \begin{table}
    \centering
    \begin{tabular}{lllllccc}
      & $h_1$ & $h_2$ & $h_3$ & $h_4$ & $m_1$ & $m_2$ & $m_3$ \\
      $1$ & \mcoreinline{false} & \mcoreinline{false} & \mcoreinline{false} & \mcoreinline{false}
          & 7 & 5 & 1 \\
      $2$ & \mcoreinline{false} & \mcoreinline{true} & \mcoreinline{false} & \mcoreinline{true}
          & 2 & 4 & 2 \\
      $3$ & \mcoreinline{true} & \mcoreinline{false} & \mcoreinline{true} & \mcoreinline{?}
          & 3 & 6 & \mcoreinline{?} \\
      $4$ & \mcoreinline{true} & \mcoreinline{true} & \mcoreinline{true} & \mcoreinline{?}
          & 6 & 3 & \mcoreinline{?} \\
    \end{tabular}
    \caption{Columns $h_1$--$h_4$ show the four possible combinations of values
      for the four holes in the dependency graph in
      Figure~\ref{fig:bipartite-partial}, for $n=4$ and $m=3$. %
      Columns $m_1$--$m_3$ show possible costs for the three measuring points.
      The cost of a measuring point could for instance be the sum (in seconds)
      of the execution times over all invocations of the measuring point.
      The \mcoreinline{?}s indicate that we may choose any value for $h_4$
      in iteration $3$ and $4$, since we have already exhausted the
      sub-graph consisting of $h_4$ and $m_3$.
      Note that the four combinations listed in the table are not unique. For
      instance, we may shift the values in the $h_1$ column one step to produce
      another table. However, there are never more than four necessary
      combinations to evaluate.}\label{tab:configurations}
  \end{table}
  \qed{}
\end{example}

We define the \emph{reduced search space size} given a dependency graph
$(H,M,E)$ as: $ \max\limits_{m\in M}\prod\limits_{(h,m)\in E} |\text{dom}(h)| $,
where $\text{dom}(h)$ denotes the domain, that is, the set of possible values,
for a hole $h$.
Applying this formula to Example~\ref{ex:dep-partial} gives $2^2=4$
configurations, as expected.g

\subsection{Choosing the Optimal Configuration}\label{sec:choosing-optimal}

\newcommand{\ResultTable}[0]{T}
\newcommand{\ConfigMatrix}[0]{C}
\newcommand{\ObservationMatrix}[0]{O}
\newcommand{\Measures}[0]{K}

This section considers how to choose the optimal configuration, given an
objective value to be minimized and the observed results of a set of hole value
combinations.
Specifically, we assume that we have:
% \begin{itemize}
a dependency graph $G=(\HolesSet{},\MeasSet{},E)$;
a configuration matrix $\ConfigMatrix{}$ of dimension $r \times
  |\HolesSet{}|$, where $\ConfigMatrix[i,j]$ gives the value of the $j$th hole
  in the $i$th iteration (compare columns $h_1$--$h_4$ in
  Table~\ref{tab:configurations});
 a number of observation matrices $\ObservationMatrix_k{}$, each with
  dimension $r \times |\MeasSet{}|$, for $k$ in some set of measures
  $\Measures{}$. For the rest of this section, we assume that there is only one
  measure, namely accumulated execution time. Thus, we denote the only
  observation matrix by $O$. That is, $O[i,j]$ gives the accumulated execution
  time for the $j$th measuring point in the $i$th iteration (compare columns
  $m_1$--$m_3$ in Table~\ref{tab:configurations}).

The problem is to assign each hole in $\HolesSet{}$ to values in their domains,
such that the objective function is minimized, where the objective function is
built from the observation matrices. In this section, we assume that the
objective is to minimize the sum of the accumulated execution times for the
measuring points. However, the approaches discussed here are general enough to
handle any number of observation matrices, with some other custom objective
function.

% Total time but could be something else?
Before presenting two general approaches for solving this problem, we consider
how to solve it for the example in Table~\ref{tab:configurations}:

\begin{example}
  Consider the results for the measuring points $m_1$, $m_2$, and $m_3$ in
  Table~\ref{tab:configurations}.
  At first glance, the optimal configuration seems to be configuration 2, since
  it has the lowest total execution time, namely $8$~s, out of the four options
  (regardless of the value of $h_4$ in iteration $3$ and $4$, the total value
  will exceed $8$).
  However, the first improvement to this is that we can choose the value of
  $h_4$ independently of the values of the other holes, since $h_4$ is disjoint
  from the other holes in the dependency graph. We see that the best value for
  $h_4$ is $\mcoreinline{false}$, giving $m_3$ the execution time $1$~s.
  The second improvement is that we can choose the value of $h_1$ independently
  of the value of $h_3$. With this in mind, the best values for $h_1$, $h_2$,
  and $h_3$ is $\mcoreinline{false}$, $\mcoreinline{true}$, and
  $\mcoreinline{true}$, respectively. This gives $m_2$ and $m_3$ the execution
  times $2$~s and $3$~s, respectively.
  Thus, the optimal configuration is one that is not explicitly listed in the
  table, and has the estimated cost of $6$~s.
  \qed{}
\end{example}

\newcommand{\TableConstraint}[0]{\texttt{table}}
\newcommand{\ListConcat}[0]{\mathit{++}}

%

% \subsubsection{Explicit Approach}
The first approach is to consider each possible combination explicitly and pick
the combination giving the lowest total execution time.
As an optimization, we can consider each disjoint part (that is, each connected
component) of the dependency graph separately. In the example in
Table~\ref{tab:configurations}, this means that we create one explicit matrix
for the connected component consisting of the vertices $\{h_1,h_2,h_3,m_1,m_2\}$
and one for the connected component with vertices $\{h_4,m_4\}$.
That is, we would infer the matrix in Table~\ref{tab:m1m2} for the measuring
points $m_1$ and $m_2$, and similarly a matrix with two rows for $m_3$. From
these explicit matrices, we can directly find that the minimum expected cost is
$6$, when $h_1=\mcoreinline{false}$, $h_2=\mcoreinline{true}$,
$h_3=\mcoreinline{true}$, and $h_4=\mcoreinline{false}$.
\begin{table}
    \centering
    \begin{tabular}{llllcc}
      & $h_1$ & $h_2$ & $h_3$ & $m_1$ & $m_2$ \\
      $1$ & \mcoreinline{false} & \mcoreinline{false} & \mcoreinline{false}
          & 7 & 5 \\
      $2$ & \mcoreinline{false} & \mcoreinline{false} & \mcoreinline{true}
          & 7 & 6 \\
      $3$ & \mcoreinline{false} & \mcoreinline{true} & \mcoreinline{false}
          & 2 & 4 \\
      $4$ & \mcoreinline{false} & \mcoreinline{true} & \mcoreinline{true}
          & 2 & 3 \\
      $5$ & \mcoreinline{true} & \mcoreinline{false} & \mcoreinline{false}
          & 3 & 5 \\
      $6$ & \mcoreinline{true} & \mcoreinline{false} & \mcoreinline{true}
          & 3 & 6 \\
      $7$ & \mcoreinline{true} & \mcoreinline{true} & \mcoreinline{false}
          & 6 & 4 \\
      $8$ & \mcoreinline{true} & \mcoreinline{true} & \mcoreinline{true}
          & 6 & 3 \\
    \end{tabular}
    \caption{Explicit listing of expected results for $m_1$ and $m_2$. The
      combinations in rows $2$, $4$, $5$, and $7$ have never been run; they are
      inferred from the observations in Table~\ref{tab:configurations}, taking
      the dependency graph into account.
    }\label{tab:m1m2}
  \end{table}

  Although the complexity of the explicit approach scales exponentially with the
  number of holes, we observe in our practical evaluation that this step is not
  a bottle neck for performance of the tuning.
  However, if this were to become a practical problem in the future, this
  problem can be solved in a more efficient way by formulating it as a
  constraint optimization problem (COP)~\cite{DBLP:reference/fai/2}. There exist
  specialized constraint programming solvers (CP solvers), that are highly
  optimized for solving general COPs, for instance Gecode~\cite{gecode} and
  OR-Tools~\cite{ortools}. The problem of choosing the optimal configuration
  given a set of observations can be expressed and solved using one of these
  solvers.

  \subsection{Exploring the Reduced Search Space}\label{sec:heuristics}
  
  In the experimental evaluation of this paper, we have implemented exhaustive
  search of the \emph{reduced search space}. As an additional step in the search
  space reduction, the tuner can optionally focus on the measuring points having
  the highest execution times. These measuring points are found by executing the
  program with random configurations of the hole values a number of times, and
  finding the measuring points with the highest mean execution times. This
  optional step reduces the search space additionally in practical experiments.
  Of course, there exist other heuristic approaches for exploring the search
  space, such as tabu search or simulated annealing. Evaluating these approaches
  is outside the scope of this paper, but they can be implemented in our modular
  tuning framework. In Section~\ref{sec:implementation}, we see that modularity
  is a key concept of the Miking language.

\section{Design and Implementation}\label{sec:implementation}

We implement the methodology of programming with holes into the Miking compiler
toolchain~\cite{Broman:2019}. Figure~\ref{fig:design} shows the design of the
implementation. In this section, we first discuss the overall design of the
toolchain, and then go through the three possible flows through it: default
compilation; tuned compilation; and tuning.

\begin{figure}
  \centering
  \includegraphics[width=0.8\textwidth,scale=1.0,trim={0cm 0cm 0cm
    0cm},clip]{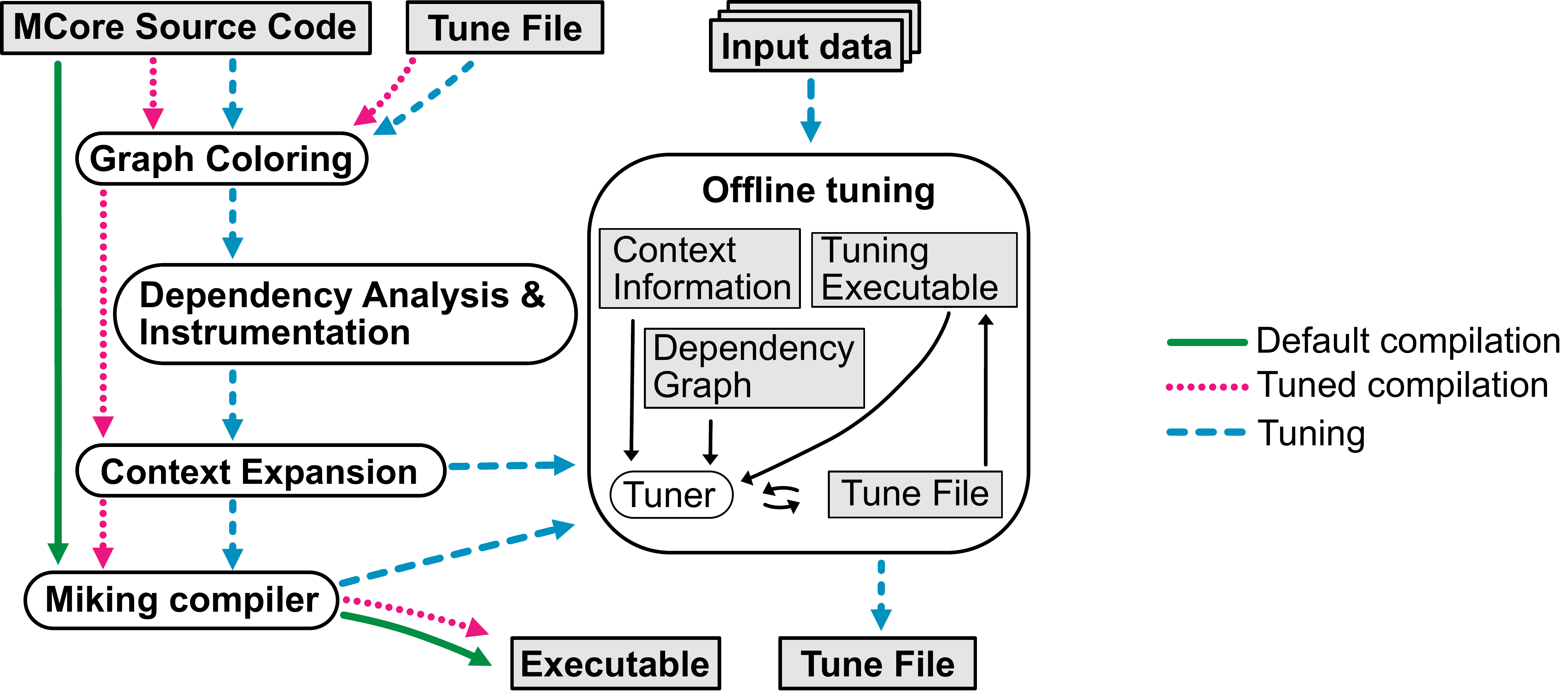}
  \caption{There are three possible flows through the Miking compiler
    toochain: %
    default compilation (green solid arrows); %
    tuned compilation (dotted magenta-colored arrows); %
    and tuning (dashed blue arrows). %
    Sections~\ref{sec:default-compilation}--\ref{sec:tuning} give a detailed
    explanation of each flow. %
    In the figure, artifacts are marked grey and has sharp corners, while
    transformations are white and rounded. }\label{fig:design}
\end{figure}

\subsection{The Miking Compiler Toolchain}
\label{sec:tool-chain}

Miking is a general language system for developing domain-specific and
general-purpose languages.
The Miking compiler is self-hosting (bootstrapped with OCaml).
The core language of the Miking system is called MCore (Miking Core) and is a
small functional language.
A key language feature of MCore is language fragments. A language fragment
defines the abstract syntax and semantics of a fragment of a programming
language. By composing several language fragments, new languages are built in a
modular way.
To extend the Miking compiler toolchain with holes, we create a new language
fragment defining the abstract syntax and semantics of holes, and compose this
fragment with the main MCore language.
The holes are transformed away before the compilation of the program.
The motivation for implementing our methodology in Miking is partly because the
system is well-designed for implementing language extensions and program
transformations, and partly because the methodology can be incorporated in any
language developed in Miking.

% cloc --force-lang-def=my_definitions.txt ~/Documents/miking-lang/miking
\newcommand{\MikingNbrFiles}[0]{$300$}
\newcommand{\MikingNbrLines}[0]{$55,000$}
\newcommand{\MikingBlankPercentage}[0]{$30$}
\newcommand{\MikingTuningNbrFiles}[0]{$18$}
\newcommand{\MikingTuningNbrLines}[0]{$5,000$}
\newcommand{\MikingTuningBlankPercentage}[0]{$30$}

The Miking toolchain consists of approximately \MikingNbrFiles~files and
\MikingNbrLines~lines of MCore code (out of which approx.
\MikingBlankPercentage$\%$ is either blank lines or comments). The contribution
of this paper is the part implementing holes (including language extensions,
program transformations, tuning, tuned compilation and dependency analysis).
This part consists of \MikingTuningNbrFiles~files and approx.
\MikingTuningNbrLines~lines of code (approx. \MikingTuningBlankPercentage$\%$
blank lines or comments).

\subsection{Default Compilation}
\label{sec:default-compilation}

The green solid path in Figure~\ref{fig:design} shows default compilation. In
this path, each hole in the program is statically replaced by its default value.
The resulting program is compiled into an executable.
Default compilation is useful during development of a program, as tuning can
take considerately longer time than default compilation.

\subsection{Tuned Compilation}
\label{sec:tuned-compilation}

The dotted magenta-colored path in Figure~\ref{fig:design} shows tuned
compilation. In this scenario, the program and the tune file are given as input
to the graph coloring, followed by context expansion
(Section~\ref{sec:transformations}).
The context expansion statically inserts the tuned values for each context into
the program.
Finally, the program is compiled into an executable.
Tuned compilation is done \emph{after} tuning has been performed, in order to
create an executable where the holes are assigned to the tuned values.
Optionally, tuned compilation can be performed automatically after tuning.

\subsection{Tuning}
\label{sec:tuning}

The blue dashed flow in Figure~\ref{fig:design} shows the offline tuning.
The program and the tune file (optional) are given as input to the graph
coloring. If the tune file is provided, then these values are considered
defaults, instead of the values provided via the \mcoreinline{default} keyword.
The graph coloring outputs (i) \emph{context information} about the holes, which
is used in the offline tuning, and in later transformation stages, and (ii) a
transformed program.
Next, the dependency analysis and instrumentation
(Section~\ref{sec:dependency-analysis}) computes a \emph{dependency graph},
which is also used in the offline tuning, and an instrumented program.
The last transformation stage, context expansion, replaces each hole with code
that looks up its current calling context.
The context expansion sends the context information and the dependency graph to
the offline tuning, and the transformed program to the Miking compiler.
The Miking compiler creates an executable to be used during tuning, the
(\emph{tuning executable}).
The \emph{offline tuning} takes the context information, dependency graph,
tuning executable, and a set of input data as input. The tuner maintains a
temporary tune file, which contains the current values of the holes. In each
search iteration, the tuning executable reads these values from the file, and
the tuner measures the runtime of the program on the set of input data.
When the tuning finishes, the tuner writes the best-found values to a final tune
file.

The tuner first reduces the search space using the dependency graph and then
applies dependency-aware tuning (Section~\ref{sec:dep-tuning}).
The stopping condition for the tuning is configurable by the user and is either
a maximum number of search iterations, or a timeout value.

\section{Empirical Evaluation}\label{sec:eval}

\newcommand{\Expressibility}[0]{1}
\newcommand{\Reduced}[0]{2}

This section evaluates the implementation. The purpose is to demonstrate that
the approach scales to real-world use cases, and to show that context-sensitive
holes are useful in these settings.
Specifically, we evaluate the following claims:
\begin{description}
\item[Claim~\Expressibility:] We can express implementation choices in
  real-world and non-trivial programs using context-sensitive holes.
\item[Claim~\Reduced:] Dependency analysis reduces the search space of
  real-world and non-trivial tuning problems.
\end{description}
The evaluation consists of three case studies of varying sizes and from
different domains. Two of the case studies, probabilistic programming and Miking
compiler, are real-world applications not originally written for the purpose of
this evaluation. The third case study, $k$-nearest neighbor classification, is
of smaller size, yet is a non-trivial program.
The experiments are run under Linux Ubuntu~18.04 ($64$~bits) on an Intel Xeon
Gold 6148 of $2.40$~GHz, with $20$~cores, hyperthreading enabled ($2$~threads
per core). The computer has $64$~GB RAM and a $1$~MB L2 cache.
As backend compiler for the Miking compiler, we use the OCaml compiler available
as an OPAM switch \texttt{4.12.0+domains}. At the time of writing, this is the
latest OCaml compiler with multicore support.

\subsection{$k$-Nearest Neighbor Classification}\label{sec:eval-knn}

This case study consists of a variant of the running example in
Section~\ref{sec:dependency-analysis}, $k$-NN classification.
Again, we consider the three implementation choices of the underlying
representation of sequence ($\Hseq$), parallelization of the \mcoreinline{map}
function ($\Hmap$), and choice of sorting algorithm ($\Hsort$).
We assume that the performance of these choices depends on the size of the input
data, and that we are interested in tuning the classifier for a range of
different sizes of the data set.
We believe that for small data sets, the sequence representation cons list is
more efficient than Rope, the \mcoreinline{map} function is more efficient when
run sequentially than in parallel, and that insertion sort is more efficient
than merge sort, respectively. However, we do not know the threshold values for
these choices. Therefore, we let the three base holes $\Hseq$, $\Hmap$ and
$\Hsort$ be of type \mcoreinline{IntRange}, representing the unknown threshold
values. For instance, assuming the $\Hmap$ hole is called
\mcoreinline{parThreshold}, the following: \mcoreinline{if lti (length seq)
  parThreshold then smap f seq else pmap f seq},
encodes the choice for the \mcoreinline{map} function.

We assume we are interested in data sets of sizes $10^3$--$10^5$ points. We set
the minimum and maximum values of the holes accordingly to
$\mcoreinline{min}=10^3$ and \mcoreinline{max} slightly higher than $10^5$, say
$\mcoreinline{max}=101,000$, respectively. That is, the \mcoreinline{min} value
corresponds to making the first choice (e.g., sequential \mcoreinline{map}) for
all input sizes, while the \mcoreinline{max} value corresponds to making the
second choice (e.g., parallel \mcoreinline{map}) for all input sizes.

We generate $6$~random sets of data points with dimension $3$, in sizes in the
range of interest: $10^3, 2\cdot 10^4, 4\cdot 10^4, 6\cdot 10^4, 8\cdot 10^4,
10^5$. We use a step size of $2\cdot 10^4$ when tuning, so that hole values with
this interval are considered.

The dependency analysis results are that the search space size is reduced by
approx.~$83\%$ (from $216$ to $36$ configurations). The best found configuration
for the threshold values were $21000$, $1000$, and $101000$ for $\Hmap$,
$\Hsort$, and $\Hseq$, respectively. That is, all input sizes except the
smallest runs \mcoreinline{map} in parallel, all input sizes use merge sort, and
all input sizes use cons lists as sequence representation.
Table~\ref{tab:knn} presents the execution time results of the tuned program
compared to the worst configuration. We see that the tuning gives
between~$3$--$1500$ speedup of the program.

We note that for this case study, allowing for tail-call optimization in the
instrumented program (see Section~\ref{sec:dep-instrumentation}) is of utmost
importance. The sorting functions are tail recursive, so an instrumented
program without tail-call optimizations gives non-representative execution
times, or even stack overflow for large enough input sizes.
The total time for the tuning is approx. $3.5$ hours, and the static analysis
takes less than $100$ ms.

\begin{figure}
   \begin{subfigure}[t]{.45\textwidth}
      \centering
      \begin{tabular}{lll}
    Input size & Execution time & Speedup \\
    \midrule
    \input{experiments/knn/knn.tex}
      \end{tabular}
        \caption{Tuning results for the $k$-NN classifier. The measurements are done
    on different data sets than were used during tuning.
    The 'Input size' column shows the size of each data set.
    The 'Execution time' column shows the execution time in seconds for the
    tuned program for a given input size.
    The 'Speedup' column presents the speedup of the tuned program compared to
    the worst configuration. The worst configuration for each input size is
    found by measuring the execution time of the program when setting the
    threshold so that it is \emph{below} respectively \emph{above} the given
    input size, for each of the three holes. That is, in total there are $8$
    ($=2^3$) candidate configurations for each input size.
  }\label{tab:knn}
\end{subfigure}\hfill
   \begin{subfigure}[t]{.45\textwidth}
     \centering
       \begin{tabular}{l|ll}
      & Sequential & Parallel (worst) \\
    \midrule
    Rope & 16.8 & 1.96 \\
    List & 18.4 & 1.94 \\
       \end{tabular}
       \caption{Speedup of the tuned probabilistic program. The
    numbers show the speedup of the
    best found configuration (Rope and
    parallel \mcoreinline{map} using a chunk size of 610 elements), compared to
    the other configurations.
    The 'Rope' and 'List' rows are using the respective sequence representation.
    The 'Sequential' column uses sequential \mcoreinline{map}, and the
    'Parallel (worst)' column uses parallel
    \mcoreinline{map}
    with the chunk size giving the worst execution time. For both Rope and List, the worst
    choice for the chunk size is $10$ elements.
    The execution time of each program is measured $10$ times, and the speedup
    is calculated as the ratio of the mean execution time of the program divided
    by the mean of the baseline. The execution time of the baseline (the tuned
    program) is $7.1 \pm 0.06$ seconds (mean and standard deviation over $10$
    runs). }\label{tab:prob}
\end{subfigure}
\caption{Results for the $k$-NN and the probabilistic programming case studies.}
\end{figure}

\subsection{Probabilistic Programming}\label{sec:eval-prob}

This case study considers a probabilistic programming framework developed consisting of approx. $150$ files and $1,500$ lines of
MCore code. Note that the majority of this code is the standard MCore code of
the general purpose program and that the probabilistic programming parts
consists of a minimal extension.
We focus on the inference backend of the framework, using the
importance sampling inference method. The inference is a core part of the
framework, and is used when solving any probabilistic programming model.
We tune the \emph{underlying sequence representations} and the \emph{map
  function} within the inference backend. The sequence representation is either
cons list or Rope. The map function chooses between a sequential or parallel
implementation.
In addition, we tune the \emph{chunk size} of the parallel implementation (see
Example~\ref{example-map3}).

We use a simple probabilistic model representing the outcome of tossing a fair
coin. The model makes $10,000$ observations of a coin flip from a Bernoulli
distribution, and infers the posterior distribution, given a Beta prior
distribution.
We expect that the choices the tuner makes are valid for a given model and
number of particles used in the inference algorithm, because these two factors
are likely to influence the execution time of the \mcoreinline{map} function.
Once a given model is tuned, however, it does not need to be re-tuned for other
sets of observed data, as long as the number of observations is the same.

We tune the model using $30,000$ particles for the inference algorithm. The
tuner chooses to use Rope as sequence representation in combination with
parallel map with a chunk size of $610$ elements.
Table~\ref{tab:prob} shows the speedup of the best found configuration compared
to the others. For instance, we see that we get a speedup of $1.96$ when using a
chunk size of $610$ for Rope compared to using the worst chunk size ($10$).
The total tuning time for the program is approx. $6$~minutes.

\subsection{Miking Compiler}\label{sec:eval-miking}

This case study considers the bootstrapping compiler, a subset of the Miking
compiler toolchain. The purpose is to test the dependency for a problem of
larger scale.
For each sequence used within the compiler, we express the choice of which
underlying representation to use (Rope or list) using a context-sensitive hole.
By default, the compiler uses Rope. Because the main use of sequences within the
compiler consists of string manipulation, which is very efficient using Rope, we
do not believe there is much to gain from using lists. However, the purpose of
this experiment is not to improve the execution time of the compiler, but rather
to show search space reduction.
After the context expansion, there are in total $2,309$ holes. That is, the size
of the original search space is $2^{2309}$.
After applying dependency analysis, the search space is reduced to $2^{924}$.
By filtering out all measuring points that have a mean execution time of less
then $10$~ms, the search space is further reduced to $2^{816}$.
The total time of the static analysis is approx.~$16$ minutes, which is
considerably higher than for the previous case studies, due to the size of this
program.

When performing this case study, we choose to disable the feature of the
instrumentation that allows for tail-call optimization, because of an identified
problem with this feature.
We only observe this problem for this large-scale program; for the other case
studies the correctness of the instrumentation is validated manually and by
assertions within the instrumented code.

\subsection{Discussion}

This section relates the claims with the results from the case studies, and
discusses correctness of possible hole values.

This evaluation considers two claims and three case studies.
Claim~\Expressibility, expressibility of implementation choices in real-world
and non-trivial programs, is shown in all three case studies. Using holes, we
can encode the automatic selection of algorithms and data structures, as well as
parallelization choices. We can also encode dependencies on data size in the
program, using threshold values.
We address Claim~\Reduced{} in the $k$-NN classification and Miking compiler
case studies. In both these cases, the search spaces are considerably reduced.
We observe, especially from the Miking compiler case study, that a possible area
for improvement in the dependency analysis is the call graph analysis step
(Section~\ref{sec:call-graph-after-cfa}). The reason is that for a large
program, dependencies from \emph{potential} executions nested measuring points
within branches in match expressions quickly accumulates, giving a quickly
growing search space. By taking into account that the execution of the nested
points are only \emph{conditionally} dependent on the condition of the match
expressions, we can reduce the search space further.

An essential and challenging aspect when programming in general is the
functional correctness of the program.
When programming with holes, this aspect can become even more challenging, as
combinations of hole values form a (sometimes complicated) set of possible
programs.
The typical software engineering approach for increasing confidence of
correctness is to use testing.
As it turns out, testing can also aid us in the case of programming with holes.
The MCore language has built-in support for tests (via the language construct
\mcoreinline{utest}), and these are stripped away unless we provide the
\texttt{--test} flag. By providing the \texttt{--test} flag when invoking the
tuning stage, the tuner will run the tests during tuning, using the currently
evaluated hole values. The result is a slight degradation in tuning time but no
overhead in the final tuned binary.
As a practical example, we use \mcoreinline{utest}s in the $k$-NN case study in
this evaluation in order to ensure that the classifier indeed chooses the
correct class for some test data sets.

\section{Related Work}\label{sec:related}

This section discusses related work within auto-tuners, by partitioning them
into domain-specific and generic tuners. We also discuss work using static
analysis within auto tuning.

Many successful auto-tuners target domain-specific problems.
SPIRAL~\cite{spiral-overview-18} is a tuning framework within the digital signal
processing domain, ATLAS~\cite{atlas} tunes libraries for linear algebra,
FFTW~\cite{fftw} targets fast Fourier transforms,
PetaBrikcs~\cite{petabricks-09} focuses on algorithmic choice and composition,
and the work by~\cite{sorting-04} automatically chooses the best sorting
algorithm for a given input list.
Moreover, within the area of compiler optimizations, a popular research field is
auto-tuning the selection and phase-ordering of compiler optimization
passes~\cite{survey-compiler-autotunings-using-ml}.
On the one hand, a natural drawback with a domain-specific tuner is that it is
not applicable outside of its problem scope, while generic tuners (such as our
framework) can be applied to a wider range of tuning problems.
On the other hand, the main strength of domain-specific tuners is that they can
use knowledge about the problem in order to reduce the search space. For
instance, SPIRAL applies a dynamic programming approach that incrementally
builds solutions from smaller sub-problems, exploiting the recursive structure
of transform algorithms.
Similarly, PetaBricks also applies dynamic programming as a bottom-up approach
for algorithmic composition, and the authors of \cite{sorting-04} include the
properties of the lists being sorted (lengths and data distribution) in the
tuning.
Such domain-dependent approaches are not currently applied in our framework,
because the tuner has no deep knowledge about the underlying problem.
We are therefore limited to generic search strategies.
An interesting research problem is investigating how problem-specific
information can be incorporated into our methodology, either from the user, or
from compiler analyses, or both.
Potentially, such information can speed up the tuning when targeting particular
problems, while not limiting the generalizability of our approach.

Among the generic tuners, CLTune~\cite{cltune} is designed for tuning the
execution time of OpenCL kernels, and supports both offline and online tuning.
OpenTuner~\cite{open-tuner} allows user-defined search-strategies and objective
functions (such as execution time, accuracy, or code size).
ATF~\cite{atf} also supports user-defined search strategies and objectives and
additionally supports pair-wise constraints, such as expressing that the value
of a variable must be divisible by another variable.
The HyperMapper~\cite{hypermapper} framework
has built-in support for multi-objective optimizations so that trade-off curves
of e.g. execution time and accuracy can be explored.
Our approach is similar to these approaches as we have a similar programming
model: defining unknown variables (holes) with a given set of values.
The difference is that we support context-sensitive holes, while previous works
perform global tuning.

There are a few previous approaches within the field of program tuning using
static analysis to speed up the tuning stage. For instance, to collect metrics
from CUDA kernels in order to suggest promising parameter settings to the auto
tuner~\cite{DBLP:conf/icpp/LimNM17}, or static analysis in combination with
empirical experiments for auto tuning tensor
transposition~\cite{DBLP:conf/ipps/WeiM14}.
To the best of our knowledge, there is no prior work in using static analysis
for analyzing dependency among decision variables.
One prior work does exploit dependency to reduce the search
space~\cite{DBLP:conf/europar/SchaeferPT09}, but their approach relies entirely
on user annotations specifying the measuring points, the independent code blocks
of the program, permutation regions, and conditional dependencies.
By contrast, our approach is completely automatic, where the user can refine the
automatic dependency analysis with certain annotations, if needed.

\section{Conclusion}\label{sec:conclusion}
In this paper, we propose a methodology for programming with holes that enables
developers to postpone design decisions and instead let an automatic tuner find
suitable solutions. Specifically, we propose two new concepts: (i)
context-sensitive holes, and (ii) dependency-aware tuning using static analysis.
The whole approach is implemented in the Miking system and evaluated on
non-trivial examples and a larger code base consisting of a bootstrapped
compiler. We contend that the proposed methodology may be useful in various
software domains, and that the developed framework can be used for further
developments of more efficient tuning approaches and heuristics.

%% Acknowledgments
\begin{acks}                            %% acks environment is optional
                                        %% contents suppressed with 'anonymous'
  %% Commands \grantsponsor{<sponsorID>}{<name>}{<url>} and
  %% \grantnum[<url>]{<sponsorID>}{<number>} should be used to
  %% acknowledge financial support and will be used by metadata
  %% extraction tools.
  % this material is based upon work supported by the
  % \grantsponsor{gs100000001}{national science
  %   foundation}{http://dx.doi.org/10.13039/100000001} under grant
  % no.~\grantnum{gs100000001}{nnnnnnn} and grant
  % no.~\grantnum{gs100000001}{mmmmmmm}.  any opinions, findings, and
  % conclusions or recommendations expressed in this material are those
  % of the author and do not necessarily reflect the views of the
      %       National Science Foundation.

  This project is financially supported by the Swedish Foundation for Strategic Research (FFL15-0032 and RIT15-0012). The research has also been carried out as part of the Vinnova Competence Center for Trustworthy Edge Computing Systems and Applications (TECoSA) at the KTH Royal Institute of Technology. We would like to thank Joakim Jald\'{e}n, Gizem \c{C}aylak, Oscar Eriksson,
  and Lars Hummelgren for valuable comments on draft versions of this paper. We
  also thank the anonymous reviewers for their detailed and
  constructive feedback.

\end{acks}

%% Bibliography
\bibliographystyle{ACM-Reference-Format}
\bibliography{paper}

\end{document}